\documentclass{pasj01}
\draft 
\Received{$\langle$reception date$\rangle$}
\Accepted{$\langle$acception date$\rangle$}
\Published{$\langle$publication date$\rangle$}
\SetRunningHead{Astronomical Society of Japan}{Usage of \texttt{pasj00.cls}}
\usepackage{subeqn}
\usepackage{subfig}
\usepackage{comment}
\usepackage{array}
\usepackage{multirow}
\usepackage{bm}
\usepackage{booktabs}
\usepackage{color}
\usepackage[font=footnotesize,labelfont=bf]{caption}
\usepackage{ulem}

\begin{document}
\title{Detailed modeling of dust distribution in the disk of HD 142527}
\author{Kang-Lou \textsc{Soon}\altaffilmark{1}$^{,*}$, Tomoyuki \textsc{Hanawa}\altaffilmark{2},
Takayuki \textsc{Muto}\altaffilmark{3},\\ Takashi \textsc{Tsukagoshi}\altaffilmark{1}, and Munetake \textsc{Momose}\altaffilmark{1,4}}
\altaffiltext{1}{College of Science, Ibaraki University, 2-1-1 Bunkyo, Mito 310-8512, Japan}
\altaffiltext{2}{Center for Frontier Science, Chiba University, 1-33 Yayoi-cho, Inage-ku, Chiba 263-8522, Japan}
\altaffiltext{3}{Division of Liberal Arts, Kogakuin University, 1-24-2 Nishi-Shinjuku, Shinjuku-ku, Tokyo 163-8677, Japan}
\altaffiltext{4}{Visiting Professor, National Astronomical Observatory of Japan, 2-21-1 Osawa, Mitaka, Tokyo, 181-8588, Japan}
\email{14nd402n@vc.ibaraki.ac.jp}
\maketitle
\begin{abstract}
We investigate the dust distribution in the crescent 
disk around HD 142527 based on the continuum emission
at $890 \mathrm{\ \mu m}$ obtained by ALMA Cycle 0. 
The map is divided into $18$ azimuthal sectors,
and the radial intensity profile in each sector 
is reproduced with a 2D disk model. 
Our model takes account of scattering and inclination
of the disk as well as the azimuthal dependence in intensity.
When the dust is assumed to have the conventional composition 
and maximum size of $1\ \mathrm{mm}$, 
the northwestern region ($PA=291^{\circ}-351^{\circ}$) cannot be reproduced.
This is because
the model intensity gets insensitive to the
increase in surface density due to heavy self-scattering,
reaching its ceiling much lower
than the observed intensity.
The ceiling depends on the position angle.
When the scattering opacity is reduced 
by a factor of $10$, the intensity distribution is 
reproduced successfully in all the sectors 
including those in the northwestern region. 
The best fit model parameters depend 
little on the scattering opacity in the southern region 
where the disk is optically thin.   
The contrast of dust surface density
along $PA$ is derived to be about $40$, much smaller than the value 
for the cases of conventional opacities ($70-130$).
These results strongly suggest that the albedo is lower than considered
by some reasons at least in the northwestern region.
\end{abstract}

\section{Introduction}
\label{introduction}
Transitional disks are believed to be in 
an evolutionary stage
from a gas-rich, primordial phase to a gas-poor, debris phase.
Unlike the primordial disks with large emission excess
between $2\ \mu\mathrm{m}$ and $25\ \mu\mathrm{m}$,
their spectral energy distributions show a dip at around $5\ \mu\mathrm{m}$,
signifying little or no emission excess in this near-infrared regime 
\citep{strom89,skrutskie90,muzerolle10,andrews11,cieza11,espaillat14}.
From the early modeling \citep{calvet02,calvet05} and
the recent high-angular resolution imaging 
at infrared and (sub-)millimeter wavelengths,
the dip is indicative 
of a dust cavity in the 
disk's inner region
\citep{brown09,grady15}.
\par
The continuum emission
at millimeter and sub-millimeter wavelengths
from the transitional disks
is
often observed to be azimuthally asymmetric
\citep{isella13,marel13,perez14}.
The asymmetry is considered to be the consequence of 
dust accumulation in regions of higher gas pressure.
The formation mechanism of such a pressure maximum is still under debate,
but it includes a large-scale vortex \citep{lyralin13,zhu14},
the perturbation by an unseen planet in the inner hole \citep{birnstiel13},
and a fast gravitational global ($m=1$) mode in the gas disk
\citep{mittalchiang15,baruteau16}.
In order to examine the validity of these proposed mechanisms,
further information about the disk structure
should be provided based on observational results.
\par
HD 142527 is a young Herbig star
of spectral type F7III \citep{vandenancker98},
surrounded by a transitional disk.
Considering its association with Sco OB2,
the distance to the star is assumed to be $140\ \mathrm{pc}$.
The mass, radius, and the effective temperature of the star
are $2.2\ M_\odot$, $3.8\ R_\odot$, and $6250\ \mathrm{K}$
\citep{fukagawa06,verhoeff11,mendigutia14}. 
A companion of $0.1\ M_\odot - 0.4\ M_\odot$ is found 
at about $13\ \mathrm{au}$ from the central star
\citep{biller12,close14,rodigas14}.
The disk structure of HD 142527 
has been studied at various wavelengths
\citep{fukagawa06,fujiwara06}.
Sub-millimeter observations yielded the outer disk 
as a crescent structure with
a cavity of the size of approximately $150\ \mathrm{au}$ in radius.
The continuum emission
in the northern region of the outer disk is brighter, thus more dust-concentrated,
than in
the southern region
\citep{casassus13}.
On the other hand, the gas distributions 
revealed by the CO rotational lines
are moderately symmetric with respect to the disk rotation axis.
From the gas kinematics traced by $^{13}$CO $(J=3-2)$, 
the inclination of the disk axis to the line of sight is derived to be $27^\circ$
\citep{fukagawa13}.
A $34\ \mathrm{GHz}$ dust clump 
is seen in the northern region of the disk,
strongly suggesting the presence of centimeter-size dust grains
\citep{casassus15}.  
\par
\citet{muto15} carried out detailed modeling of
dust and gas radial distribution in 
$PA=11^\circ - 31^\circ$ and $PA=211^\circ - 231^\circ$,
the brightest and the faintest position angle ($PA$) sectors 
in the continuum emission.
They try to reproduce the radial profiles of the dust continuum emission at $\lambda = 890\ \mu\mathrm{m}$
and line emissions of $^{13}$CO $(J=3-2)$ and C$^{18}$O $(J=3-2)$ 
in the two sectors.
Their results are consistent with dust accumulation in pressure maxima;
while the contrast of dust surface density along $PA$ is $\sim70$,
that in gas surface density is only $\sim 4$.
In order to reveal the whole dust distribution that may be the key to 
understanding
the formation mechanism of the crescent structure,
we extend the modeling of dust continuum emission
in
all the $PA$ directions.
We use the same modeling method adopted by \citet{muto15}.
\par
This paper is organized as follows.
In Section \ref{observation} we summarize the observational data used in this study. 
We introduce our modeling procedure in Section \ref{modeling},
and present the results in Section \ref{results}.
Section \ref{discussions} is reserved for discussions.

\section{Observational data}
\label{observation}
We use the continuum data of HD 142527 
taken by ALMA over the period from 2012 June to August (ADS/JAO.ALMA\#2011.0.00318.S).
Six scheduling blocks were carried out by using the Extended Array Configuration in Cycle 0
consisting of $20$ to $26$ $12$-meter antennas,
forming a maximum baseline of about $480\ \mathrm{m}$.
The correlator was configured to store dual polarizations in four separate spectral windows.
The central frequencies for the windows are $330.588$, $329.331$, $342.883$, and $342.400\ \mathrm{GHz}$;
each window has a $469\ \mathrm{MHz}$ bandwidth over $3840$ channels.
Aggregating all the line-free channels in the spectral windows,
we obtain a $1.8\ \mathrm{GHz}$ bandwidth 
for the continuum data at $336\ \mathrm{GHz}$, or $\lambda_\mathrm{obs}=890\ \mu\mathrm{m}$.
The on-source integration time is
three hours
after flagging aberrant data.
\begin{figure}
\begin{center}
\resizebox{0.45\textwidth}{!}{\input{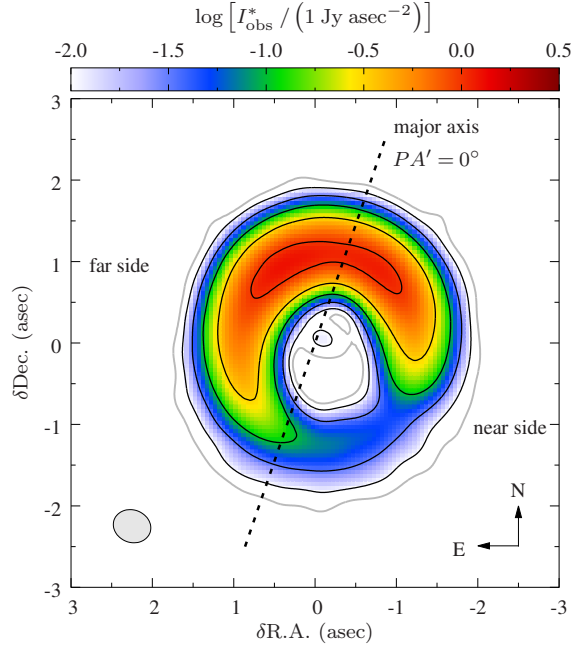}}
\caption{Image of the continuum emission at $\lambda_\mathrm{obs}=890\ \mu\mathrm{m}$
taken with ALMA.
Black contours denote the intensity 
$I^*_{\mathrm{obs}}\ \left(\mathrm{Jy}\ \mathrm{asec^{-2}}\right)$ at $\log \left [I^*_{\mathrm{obs}}\right]=-2.0$, $-1.5$, $-1.0$, $-0.5$, $0.0$.
The gray contours denote the intensity at 
$5\ \sigma$ level ($1\ \sigma=0.61\ \mathrm{mJy\ asec^{-2}}$).
The synthesized beam is shown in the bottom left corner.
The far side is in the northeast while the near side is in the southwest.}
\label{figcontinuummap}
\end{center}
\end{figure}
\begin{table}[htbp!]
\centering
\caption{
Gaussian-fitted results for the observed radial intensity profiles.}
\label{tablegaussfitobs}
\captionsetup[subfloat]{position=top}
\begin{tabular}{cccc}
\toprule
\multicolumn{1}{c}{\multirow{2}{*}{$PA'$}}     &  $I^*_{0,\mathrm{obs}}$     &   $w^*_{0,\mathrm{obs}}$   &   $r^*_{0,\mathrm{obs}}$  \\
\multicolumn{1}{c}{} &   $(\mathrm{Jy\ asec^{-2}})$  &   $(\mathrm{au})$   &   $(\mathrm{au})$  \\
\cmidrule(){1-1} \cmidrule(l){2-4}
$10^{\circ}-30^{\circ}$    &  $1.16$  &  $50.9$  &  $158.4$ \\
$30^{\circ}-50^{\circ}$    &  $1.25$  &  $51.0$  &  $152.3$ \\
$50^{\circ}-70^{\circ}$    &  $1.19$  &  $50.2$  &  $148.6$ \\
$70^{\circ}-90^{\circ}$    &  $0.94$  &  $49.0$  &  $143.1$ \\
$\phantom{1}90^{\circ}-110^{\circ}$   &  $0.72$  &  $47.0$  &  $142.6$ \\
$110^{\circ}-130^{\circ}$  &  $0.55$  &  $45.9$  &  $145.2$ \\
$130^{\circ}-150^{\circ}$  &  $0.39$  &  $45.5$  &  $150.6$ \\
$150^{\circ}-170^{\circ}$  &  $0.22$  &  $45.6$  &  $159.7$ \\
$170^{\circ}-190^{\circ}$  &  $0.11$  &  $46.5$  &  $169.5$ \\
$190^{\circ}-210^{\circ}$  &  $0.07$  &  $47.8$  &  $179.2$ \\
$210^{\circ}-230^{\circ}$  &  $0.06$  &  $50.5$  &  $178.3$ \\
$230^{\circ}-250^{\circ}$  &  $0.05$  &  $51.6$  &  $174.1$ \\
$250^{\circ}-270^{\circ}$  &  $0.13$  &  $47.9$  &  $161.5$ \\
$270^{\circ}-290^{\circ}$  &  $0.34$  &  $48.7$  &  $158.9$ \\
$290^{\circ}-310^{\circ}$  &  $0.62$  &  $48.2$  &  $157.9$ \\
$310^{\circ}-330^{\circ}$  &  $0.96$  &  $48.5$  &  $158.6$ \\
$330^{\circ}-350^{\circ}$  &  $1.16$  &  $48.3$  &  $160.4$ \\
$350^{\circ}-10^{\circ}\phantom{3}$   &  $1.13$  &  $49.3$  &  $159.9$ \\
\bottomrule
\end{tabular}
\end{table}
\par
Calibration and data reduction are made with the Common Astronomy Software Applications package version 3.4.
We use Briggs weighting with a robust parameter of $0.5$ during the self-calibration,
in order to recover the weak and extended components of the dust emission \citep{muto15}.
The synthesized beam (FWHM) is $0\farcs 40 \times 0\farcs 47$ ($PA=59 \fdg 9$), or 
$56\ \mathrm{au} \times 66\ \mathrm{au}$ at the distance of HD 142527.
The RMS noise level is $0.61\ \mathrm{mJy\ asec^{-2}}$.
Figure \ref{figcontinuummap} shows the self-calibrated image of the disk surrounding HD 142527.
\par
As the groundwork in our modeling, 
we create radial intensity profiles of the continuum emission 
for $18$ azimuthal sectors with a radial bin width of $14\ \mathrm{au}$.
We fit these discrete averaged 
intensities
in every sector by using a Gaussian function,
\begin{eqnarray}
\label{eqngaussobs}  
I^*_{\mathrm{obs}}&&\left(r^*_{\mathrm{obs}}\right)=I^*_{0,\mathrm{obs}}\ \exp \left [-{\left (\frac{r^*_{\mathrm{obs}}-r^*_{0,\mathrm{obs}}}{w^*_{0,\mathrm{obs}}} \right )}^2 \right ],
\end{eqnarray}
where $r^*_{\mathrm{obs}}$
denotes the projected radius from the star,
$I^*_{0,\mathrm{obs}}$ 
the peak intensity at $r^*_{0,\mathrm{obs}}$,
and $w^*_{0,\mathrm{obs}}$ the projected width of the radial profile.
The results of Gaussian-fitting are listed in table \ref{tablegaussfitobs}.
In this paper, we use $PA'$, the angle measured from the disk major axis,
to denote the azimuthal position of the disk instead of $PA$,
i.e., $PA'=PA+19^\circ$. 
Equation (\ref{eqngaussobs}) is a function of $PA'$,
but we have omitted it for convenience.

\section{Modeling procedure}
\label{modeling}
\begin{figure}[h!]
\begin{center}
\input{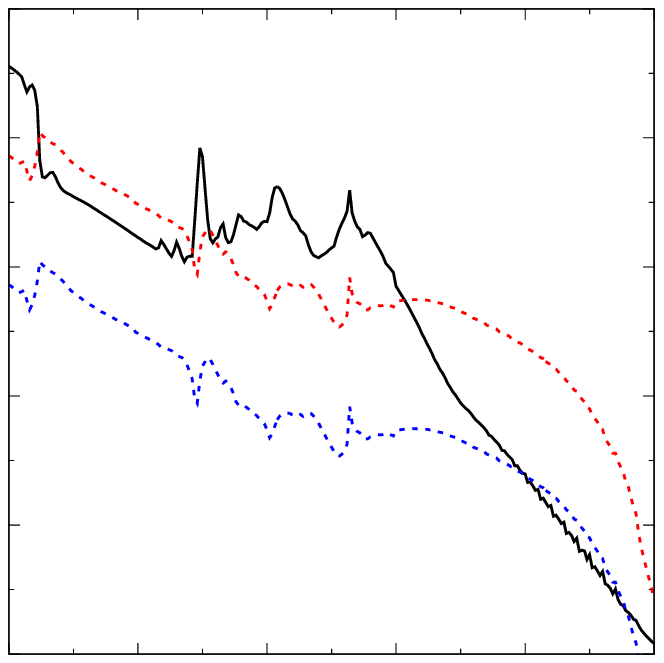}
\caption{Conventional values of the absorption opacity (black solid line) 
and the scattering opacity (red dashed line) per unit dust mass,
for compact dust grains following distribution $a^{-3.5}$, with $a_\mathrm{max}=1\ \mathrm{mm}$. 
The blue dashed line indicates the scattering opacity reduced to $10\%$ from their conventional values.}
\label{figopacity}
\end{center}
\end{figure}
\begin{figure*}[h!]
\begin{center}
\subfloat[]{\resizebox{0.45\textwidth}{!}{\input{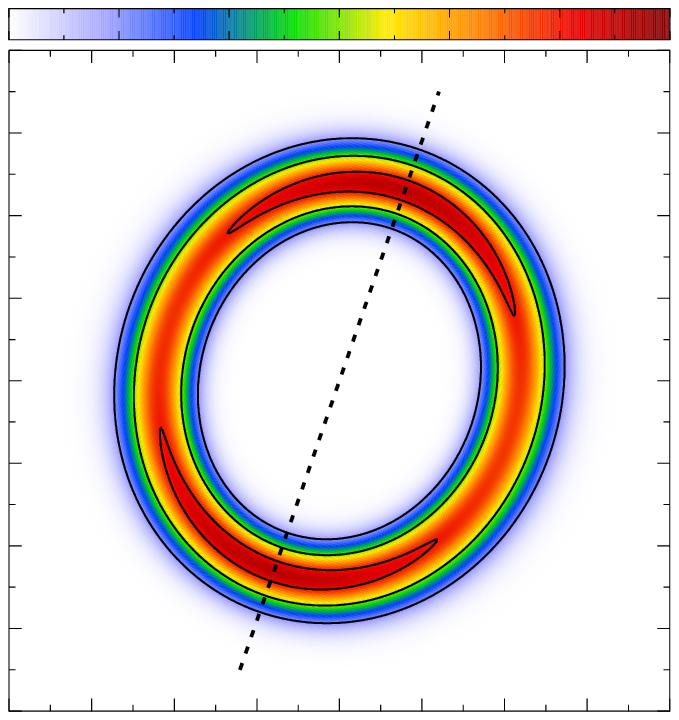}}}
\hspace{10mm}
\subfloat[]{\resizebox{0.45\textwidth}{!}{\input{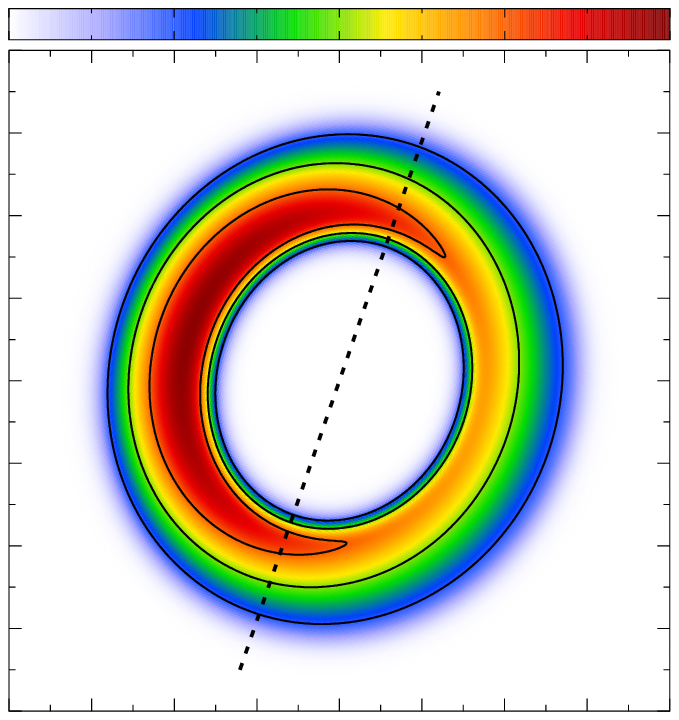}}}
\caption{Disk intensity images obtained by ray-tracing simulation.
$I'$ is the intensity before beam convolution.  
The dashed line indicates the major axis of the disk along $PA'=0^\circ$.
The simulated images are based on $w_0=27\ \mathrm{au}$ and $r_0=175\ \mathrm{au}$ models.
\textbf{(a):} $\Sigma_0 = 0.01\ \mathrm{g\ cm^{-2}}$ model. The contours are drawn at $I'=0.02$, $0.06$, $0.10 \ \mathrm{Jy\ asec^{-2}}$. 
\textbf{(b):} $\Sigma_0 = 1.00\ \mathrm{g\ cm^{-2}}$ model. The contours are drawn at $I'=0.50$, $1.00$, $1.50 \ \mathrm{Jy\ asec^{-2}}$. 
}
\label{figdiskmapbcon}
\end{center}
\end{figure*}
Our goal is to derive the dust surface density distribution for the disk
that reproduces $I^*_{0,\mathrm{obs}}$, $w^*_{0,\mathrm{obs}}$, and $r^*_{0,\mathrm{obs}}$ in each $PA'$ sector.
Since the observed intensity of the disk around HD 142527 varies 
smoothly in the azimuthal direction,
we employ a 2D axisymmetric model 
as a fiducial model to reproduce the intensity in each sector individually (Muto et al. 2015).
The scattered light of the observed 
crescent disk, however,
may be different from that of the axisymmetric model
when scattering is dominant;
the validity of this approach is discussed in Section \ref{ipeakdependence}.
\par
The dust surface density distribution in the radial ($r$) direction is assumed to be
\begin{eqnarray}  
\Sigma\left(r\right)=\Sigma_0\ \exp \left\{ -\mathrm{min} \left [{\left (\frac{r-r_{0}}{w_0} \right )}^2 ,\ 20 \right ] \right\},
\label{eqnmodelsigma}
\end{eqnarray}
where $\Sigma_0$, $w_0$, and $r_0$ 
denote the peak surface density, the radial width of the distribution,
and the peak radius, respectively.
Note that the floor $\exp{(-20)}$ is set for computational convenience. 
We determine $(\Sigma_0, w_0, r_0)$ by the following procedure:
\begin{enumerate}
\item Compute the surface density distribution, $ \Sigma (r) $, for given sets of
	$ (\Sigma _0, w _0,  r_0) $.
	The model parameters are varied in the range of
	$ -2.1 \le \log \Sigma _0~(\mathrm{g~cm^{-2}}) \le 0.1 $
	with the interval $ \Delta \log \Sigma _0 = 0.1 $,
	in $ 21~\mathrm{au} \le w_0 \le 31~\mathrm{au}$ 
	with  $ \Delta w_0  = 2~\mathrm{au}$,
	and in $159~\mathrm{au} \le r_0 \le 199~\mathrm{au}$ 
	with $ \Delta r _0  = 8~\mathrm{au}$.   
	They are extended slightly when necessary.
\item Calculate the density $\rho\left(r,z\right)$, temperature $T\left(r,z\right)$,
	and mean intensity $J_{\nu}\left(r,z\right)$ for each $(\Sigma_0, w_0, r_0)$
         by assuming the radiative and hydrostatic equilibria (Section \ref{radiationtransfer}).
\item Calculate model intensity by ray-tracing,
	and extract the radial profiles after convolving the image with the synthesized beam of the observation
	 (Section \ref{syntheticobservation}).
\item Fit the radial profiles of the model intensity 
	with the Gaussian function and obtain their $(I^*_{0}, w^*_{0}, r^*_{0})$ of the model, 
	which are compared with $(I^*_{0,\mathrm{obs}}, w^*_{0,\mathrm{obs}}, r^*_{0,\mathrm{obs}})$
	in the same $PA'$ sector (Section \ref{comparecriteria}). 
	Search for parameters $(\Sigma_0, r_0, w_0)$ that 
	meet the criteria in equation (\ref{criteria}).
	If necessary, construct a refined model by interpolating $\Sigma_0$ and $r_0$.
\end{enumerate}
In the following, we explain the procedure in detail.
\subsection{Assumptions on dust properties}
\label{dustmodel}
We assume a dust grain in the disk to be a compact sphere composed of silicate, carbonaceous compounds, and water ice;
their mass fractional abundances relative to the hydrogen mass are taken 
to be  $0.0043$, $0.0030$, and $0.0094$, respectively,
which are consistent with 
the solar elemental abundance
\citep{anders89}.
The grains have a power-law size distribution,
\begin{eqnarray}
\label{eqsizedistribution}
\frac{dn(a)}{da} \propto a^{-3.5} ,\ a_{\mathrm{max}}=1\mathrm{\ mm},
\end{eqnarray}
where $a$ and $a_{\mathrm{max}}$ denote the grain radius and the maximum grain radius, respectively.
The choice of $a_{\mathrm{max}}$ is based on the observed spectral opacity index, $\beta \sim 1$ \citep{verhoeff11}.
The absorption and scattering cross sections,
$C_\mathrm{a}(a)$ and $C_\mathrm{s}(a)$,
and the asymmetry parameter $g(a)$
\footnote{The asymmetry parameter
describes the anisotropy in scattering
and takes value within $|g| \leq 1$
(see, e.g., \cite{bh83}).
For symmetric scattering about the axis perpendicular to the 
propagation direction of incident wave, $g=0$. 
If the grain tends to scatter light toward the forward (backward) direction, 
$g$ is positive (negative).}
for every grain size in the power-law distribution
are calculated using Mie theory.
The scattering cross section and its asymmetry parameter are defined as
\begin{eqnarray}
	C_{\mathrm{s}}(a) =  \int_{4\pi} \frac{dC_{\mathrm{s}}(a)}{d\Omega} d\Omega
\end{eqnarray}
and
\begin{eqnarray}
\label{eqg}
	g(a) = \frac{1}{C_\mathrm{s}(a)}\int_{4\pi} \frac{dC_{\mathrm{s}}(a)}{d\Omega} \cos\Theta(a) d\Omega,
\end{eqnarray}
respectively,
where $dC_\mathrm{s}(a)/d\Omega$ is the differential scattering cross section
and
$\Theta(a)$ the deflection angle from the propagation direction $\Theta(a)=0^\circ$.
To take into account the effects of asymmetric scattering in the M1 model
(see Appendix \ref{appendix1} for more details),
we incorporate the asymmetry parameter
by multiplying $C_\mathrm{s}(a)$ by $[1-g(a)]$;
the light scattered away from the propagation direction 
is treated as extinction along $\Theta=0^\circ$.
In this way,
the effects of self-scattering
are considered while
keeping the computational time for radiative transfer reasonable.
Denoting the grain mass by $m(a)$,
the scattering opacity used in this study is then defined as
\begin{eqnarray}
\label{eqscattering}
\kappa_\mathrm{s} &=&  \int \left\{ \frac{C_\mathrm{s}(a)}{m(a)}  \left[ 1- g(a) \right]  \right\}  \frac{dn(a)}{da}da,
\end{eqnarray}
which is weighted by the size distribution.
Similarly, dividing $C_\mathrm{a}(a)$ by $m(a)$ and integrate over the size distribution we obtain 
the absorption opacity as 
\begin{eqnarray}
\kappa_\mathrm{a} &=&  \int \left\{  \frac{C_\mathrm{a}(a)}{m(a)}  \right\} \frac{dn(a)}{da}da.
\end{eqnarray}
The opacities are shown in figure \ref{figopacity}.
When $a_{\mathrm{max}}$ is comparable to $\lambda_{\mathrm{obs}}$, 
$\kappa_{\mathrm{s}}$ is larger than $\kappa_{\mathrm{a}}$ by a factor of about $10$;
$\kappa_{\mathrm{a}} = 2.9 \mathrm{\ cm^2\ g^{-1}}$ and $\kappa_{\mathrm{s}} = 26.2 \mathrm{\ cm^2\ g^{-1}}$
at $\lambda_{\mathrm{obs}} = 890\mathrm{\ \mu m}$
\citep{aikawa06}.
As will be described in Section \ref{scaredmodel},
we also use model where $\kappa_{\mathrm{s}}$ 
is reduced by $90\%$ from the conventional value.
\subsection{Disk models}
\label{radiationtransfer}
Applying the dust properties and the dust distribution,
we solve the radiative transfer of the disk 
by the M1 model
(\cite{gonzalez07,kanno13}; see also Appendix \ref{appendix1}).
The 2D distributions of the dust density $\rho(r,z)$, the temperature $T(r,z)$, and the mean intensity $J_\nu(r,z)$ 
are determined by the same method as Muto et al. (2015).
We consider the star as the heat source
and use $226$ colors in $0.1\ \mu\mathrm{m} \leq \lambda \leq 3.16\ \mathrm{mm}$ (spectral resolution of $\Delta\log\lambda=0.02$)
to calculate the radiative
and hydrostatic equilibria.
In the cylindrical coordinates $(r,\varphi,z)$,
the star is located at the origin, while
the disk is assumed to be axisymmetric around the $z$-axis
and reflection symmetric with respect to the midplane ($z=0$).
The computation region covers $30 \ \mathrm{au} \leq r \leq 410\ \mathrm{au}$ 
and $0 \ \mathrm{au} \leq  z \leq 120\ \mathrm{au}$
with a spatial resolution of $2\ \mathrm{au}$.
We did not take
the effects of accretion into account.
We also did not consider the shadows cast by the inner warped disk discussed by Marino et al. (2015).
\subsection{Model Image} 
\label{syntheticobservation}
The emergent intensity of the model is calculated 
by ray-tracing along the line of sight $z'$:
\begin{eqnarray}
\frac{dI'}{dz'}=-(\kappa_{\mathrm{a}}+\kappa_{\mathrm{s}}) \rho \left (I'-S_{\nu}\right),
\label{equationraytracing}
\end{eqnarray}
where the source function is given by
\begin{eqnarray}
S_{\nu}&=& \left(1-\eta \right)B_{\nu}\left(T\right)+\eta J_{\nu}.
\label{equationsourcefunction}
\end{eqnarray}
The source function is the sum of the Planck function $B_{\nu}\left(T\right)$ 
and the angular-averaged intensity $J_\nu$,
each weighted by $(1-\eta)$ and $\eta$, where
\begin{eqnarray}
\eta \equiv  \frac{\kappa_{\mathrm{s}}}{\kappa_\mathrm{a}+\kappa_\mathrm{s}} ,
\label{equationalbedo}
\end{eqnarray}
is the albedo.
We obtain intensity distribution on 
the uniform grid of $2\ \mathrm{au}$ on the sky plane
to construct the model image.
\par
Figure \ref{figdiskmapbcon} shows the intensity images obtained
by ray-tracing for two peak surface density models,
$\Sigma_0=0.01\ \mathrm{g\ cm^{-2}}$ and $1.00\ \mathrm{g\ cm^{-2}}$,
with $w_0=27\ \mathrm{au}$, $r_0=175\ \mathrm{au}$, and $i=27^{\circ}$.
The former represents an optically thin disk, while the latter 
represents
an optically thick disk.
In the following, the coordinates of the disk projected in the sky are denoted by $(r',PA')$, i.e., coordinates with primes ($'$).
The regions of $0^\circ<PA'<180^\circ$ and $180^\circ<PA'<360^\circ$
correspond to the far side and the near side, respectively.
The disk images appear slightly elongated in the major axis ($PA'=0^{\circ}$) 
because of the disk inclination, 
but they are reflection symmetry with respect to the minor axis ($PA'=90^{\circ}$).
The intensity depends on $PA'$ although the model is axisymmetric.
It is highest on the major axis in the optically thin disk (figure \ref{figdiskmapbcon}a),
but on the far side in the optically thick disk 
(figure \ref{figdiskmapbcon}b).
When the disk is optically thin,
the intensity
is roughly proportional to the amount of dust along the line of sight.
The major axis is brightest since the line of sight 
intersects the disk plane longer.  
When the disk is optically thick, 
the hot inner wall facing the star
is exposed to us
on the far side,
thus its intensity is higher than 
that on
the near side where the wall is hidden from sight.
We take 
this $PA'$ dependence into account in our analysis.
\par
The next step is the convolution of the above images with a Gaussian beam 
to mimic the observation.
Physical quantities obtained after beam convolution are indicated with asterisks ($^*$).
Since the beam is elliptical with its major axis lying along $PA'=78.9^{\circ}$,
the image no longer retains its reflection symmetry with respect to the minor axis after the beam convolution.
We then extract the radial intensity profiles from the convolved image
and fit them with Gaussian function to obtain
 $I^*_{0}$, $w^*_{0}$, and $r^*_{0}$.
\subsection{Comparison between observation and simulation}
\label{comparecriteria}
For each sector, a parameter cube $(I^*_{0}, w^*_{0}, r^*_{0})$ associated with $(\Sigma_{0}, w_{0}, r_{0})$ is created. 
The best fit parameters for the dust distribution 
are searched by
comparing the Gaussian-fitted values between the model and the observation.
The best fit model should satisfy the following criteria:
\begin{subeqnarray}
\label{criteria}
\left | \Delta I^*_{0} \right |  &\equiv&  \left | I^*_{0,\mathrm{obs}} - I^*_{0} \right | \leq 0.01\ I^*_{0,\mathrm{obs}}, \\
\left | \Delta w^*_0 \right |  &\equiv& \left |w^*_{0,\mathrm{obs}} - w^*_{0} \right | \leq 1\mathrm{\ au},\\
\left | \Delta r^*_0 \right |  &\equiv&  \left |r^*_{0,\mathrm{obs}} - r^*_{0} \right | \leq 1\mathrm{\ au}.
\end{subeqnarray}
In order to discern the trend in the $PA'$ direction,
criteria better than the beam resolution are set.
These criteria are different from those of \citet{muto15},
who placed a $\pm 10\%$ tolerance for all the criteria above.      
\par
To facilitate the parameter search for
the dust distribution and to save the computational time,
we interpolate models to
complement our sample on $\Sigma_0$ and $r_0$.
The interpolation
among four adjacent
$(\Sigma_0, w_0, r_0)$ models at $(r,z)$
is carried out by the following weighted average:
\begin{eqnarray}
A   &&\left(r,z; \Sigma_{0,\mathrm{i}}, w_0 , r_{0,\mathrm{i}}\right)  = \nonumber \\
     &&\sum\limits_{n=1}^{2} l_n \left[  \sum\limits_{m=1}^{2} k_m A \left( r + r_{0,m} - r_{0,\mathrm{i}} , z ; \Sigma_{0,n}, w_0 , r_{0,m} \right)\right],
\end{eqnarray}
where $A$ represents $\rho$, $T$, or $J_\nu$.
The weighting factors $f_m$ and $g_n$ 
are defined as
\begin{eqnarray}
k_1 &=& 1 - \left |   \frac{r_{0,\mathrm{i}} - r_{0,1}}{r_{0,2}-r_{0,1}}     \right |    \nonumber \\
k_2 &=&  1 - k_1   \nonumber \\
l_1 &=& 1 - \left |   \frac{\Sigma_{0,\mathrm{i}} - \Sigma_{0,1}}{\Sigma_{0,2}-\Sigma_{0,1}}     \right |    \nonumber \\
l_2 &=& 1 - l_1 \nonumber
\end{eqnarray}
The subscripts ``$1$'' and ``$2$'' specify
the two existing models,
while the subscript ``$\mathrm{i}$''  
specifies the interpolated product.
We do not perform interpolation on $w_0$ because its $2$-$\mathrm{au}$ interval in
the parameter space is sufficiently refined to fulfill equation (\ref{criteria}b).

\section{Results}
\label{results}
\begin{table*}[h!]
\centering
\caption{Derived best fit values using the conventional model for dust opacity.}
\label{tablebestfitori}
\captionsetup[subfloat]{position=top}
\begin{tabular}{cccccccc}
\toprule
\multicolumn{1}{c}{\multirow{2}{*}{$PA'$}}        &    $\Sigma_0$  &   $w_{0}$   &   \multicolumn{1}{c}{$r_{0}$}  & $M_\mathrm{sec}$ & $\Delta I^*_{0}$ & $\Delta w^*_0$ & $\Delta r^*_0$  \\
\multicolumn{1}{c}{}        &    $\left(\mathrm{g\ cm^{-2}}\right)$  &   $\left(\mathrm{au}\right)$   &   \multicolumn{1}{c}{$\left(\mathrm{au}\right)$} 
& $\left(M_{\odot}\right)$ & $\left( \times I^*_{0,\mathrm{obs}}\right)$& $\left(\mathrm{au}\right)$ & $\left(\mathrm{au}\right)$ \\
\cmidrule(){1-1} \cmidrule(l){2-5} \cmidrule(l){6-8}
$ 10^{\circ}-  30^{\circ}$  &  $7.89 \times 10^{-1}$  &  $23$  &  $175$ & $2.1 \times 10^{-4}$  & $-1.3\times10^{-5}$ & $+0.8$ & $-0.5$\\
$ 30^{\circ}-  50^{\circ}$  &  $1.12                     $  &  $21$  &  $173$ & $2.6 \times 10^{-4}$  & $-2.9\times10^{-5}$ & $+0.3$ & $-0.3$\\
$ 50^{\circ}-  70^{\circ}$  &  $7.28 \times 10^{-1}$  &  $21$  &  $173$ & $1.5 \times 10^{-4}$  & $-1.0\times10^{-5}$ & $-0.3$ & $-0.5$\\
$ 70^{\circ}-  90^{\circ}$  &  $3.07 \times 10^{-1}$  &  $21$  &  $167$ & $5.9 \times 10^{-5}$  & $-2.5\times10^{-5}$ & $+0.1$ & $-0.3$\\
$\phantom{1}90^{\circ}-110^{\circ}$  &  $2.00 \times 10^{-1}$  &  $21$  &  $165$ & $3.8 \times 10^{-5}$  & $-4.0\times10^{-5}$ & $-0.4$ & $+0.8$\\
$110^{\circ}-130^{\circ}$  &  $1.33 \times 10^{-1}$  &  $23$  &  $165$ & $2.7 \times 10^{-5}$  & $-1.6\times10^{-5}$ & $-0.2$ & $-0.1$\\
$130^{\circ}-150^{\circ}$  &  $7.83 \times 10^{-2}$  &  $23$  &  $165$ & $1.9 \times 10^{-5}$  & $-4.0\times10^{-5}$ & $+0.4$ & $+0.6$\\
$150^{\circ}-170^{\circ}$  &  $3.61 \times 10^{-2}$  &  $27$  &  $171$ & $1.1 \times 10^{-5}$  & $-1.1\times10^{-5}$ & $+0.2$ & $-0.8$\\
$170^{\circ}-190^{\circ}$  &  $1.61 \times 10^{-2}$  &  $29$  &  $177$ & $5.7 \times 10^{-6}$  & $-4.3\times10^{-5}$ & $+0.5$ & $-0.0$\\
$190^{\circ}-210^{\circ}$  &  $1.10 \times 10^{-2}$  &  $29$  &  $189$ & $4.1 \times 10^{-6}$  & $-4.5\times10^{-5}$ & $+0.5$ & $-0.6$\\
$210^{\circ}-230^{\circ}$  &  $9.05 \times 10^{-3}$  &  $32$  &  $193$ & $3.5 \times 10^{-6}$  & $-1.3\times10^{-5}$ & $-0.7$ & $+0.7$\\
$230^{\circ}-250^{\circ}$  &  $8.31 \times 10^{-3}$  &  $32$  &  $195$ & $3.0 \times 10^{-6}$  & $-4.6\times10^{-5}$ & $-0.7$ & $+0.8$\\
$250^{\circ}-270^{\circ}$  &  $2.89 \times 10^{-2}$  &  $23$  &  $185$ & $6.8 \times 10^{-6}$  & $-1.7\times10^{-5}$ & $-0.4$ & $+0.1$\\
$270^{\circ}-290^{\circ}$  &  $8.60 \times 10^{-2}$  &  $23$  &  $183$ & $2.0 \times 10^{-5}$  & $-3.6\times10^{-5}$ & $+0.2$ & $-0.0$\\
$290^{\circ}-310^{\circ}$  &  $2.21 \times 10^{-1}$  &  $23$  &  $179$ & $4.9 \times 10^{-5}$  & $-1.5\times10^{-5}$ & $+0.3$ & $+0.1$\\
\cmidrule(){1-8}
\footnotemark[$*$]$310^{\circ}-330^{\circ}$  &  $1.26$  &  $21$  &  $179$ & $3.0 \times 10^{-4}$  & $(-5.7\times10^{-2})$ & $(-2.9)$ & $+0.9$ \\ 
\footnotemark[$*$]$330^{\circ}-350^{\circ}$ &   $1.26$ &   $21$ &   $177$ & $3.2 \times 10^{-4}$  & $(-1.6\times10^{-1})$ & $(-1.2)$ & $-0.4$ \\
\footnotemark[$*$]$350^{\circ}-10^{\circ}\phantom{3}$ & $1.26$ & $21$ & $173$ & $3.2 \times 10^{-4}$ & $(-3.4\times10^{-2})$ & $+0.7$ & {$+1.0$} \\    
\bottomrule
\end{tabular}
\begin{tabnote}
\begin{flushleft}
\footnotemark[$*$] $PA'=310^\circ-330^\circ$, $PA'=330^\circ-350^\circ$,
and $PA'=350^\circ-10^\circ$ are failed to be reproduced using the conventional model.
The parenthesized parameters are those that do not fulfill the criteria in equation (\ref{criteria}).
\end{flushleft}
\end{tabnote}
\end{table*}
\subsection{Derived best fit values}
\label{bestfitresults}
\begin{figure*}[htbp]
\begin{center}
\subfloat[]{\resizebox{0.45\textwidth}{!}{\input{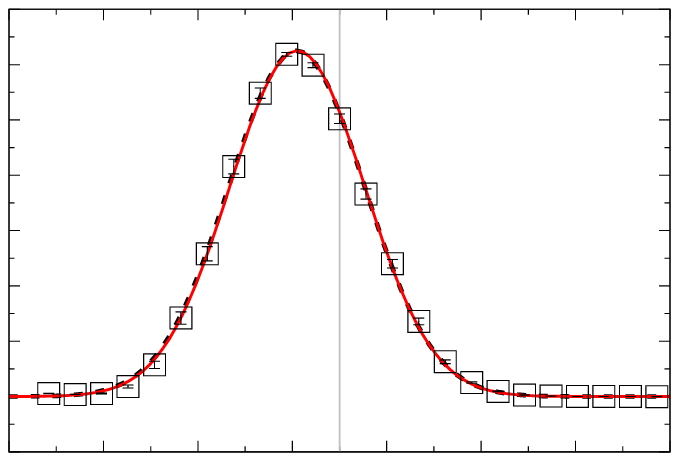}}}
\hspace{10mm}
\subfloat[]{\resizebox{0.45\textwidth}{!}{\input{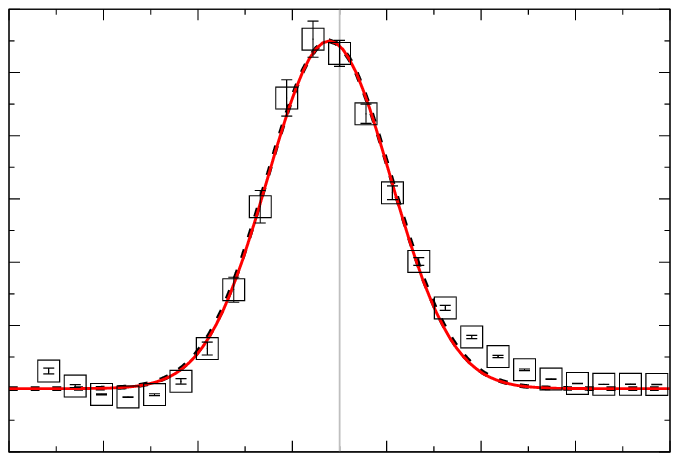}}}
\\
\subfloat[]{\resizebox{0.45\textwidth}{!}{\input{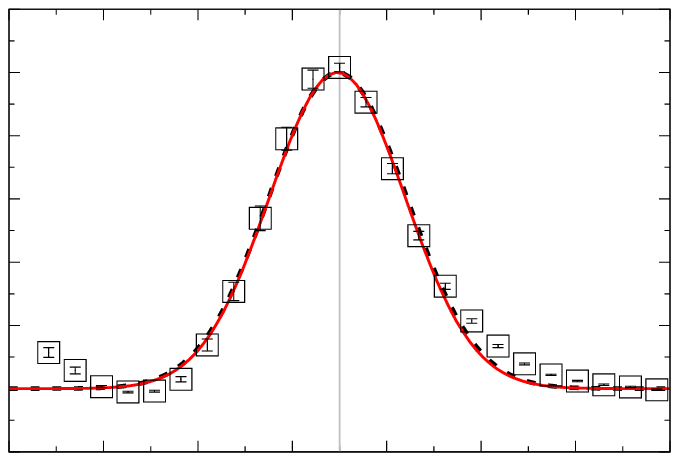}}}
\hspace{10mm}
\subfloat[]{\resizebox{0.45\textwidth}{!}{\input{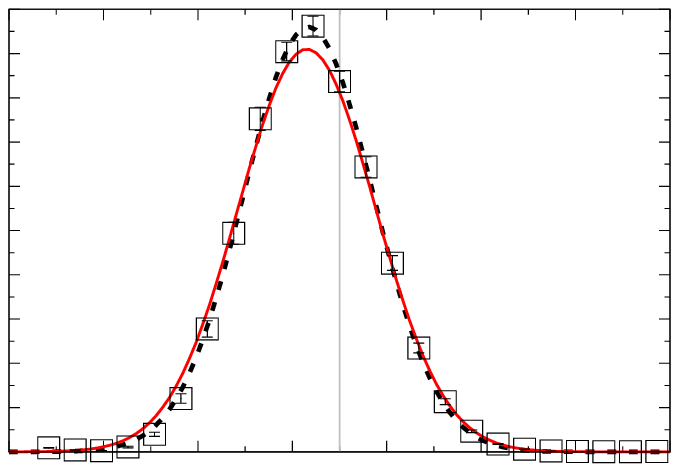}}}
\\
\subfloat[]{\resizebox{0.45\textwidth}{!}{\input{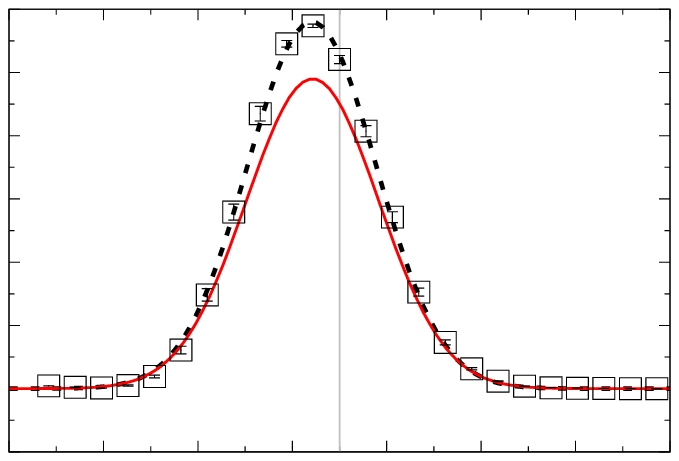}}}
\hspace{10mm}
\subfloat[]{\resizebox{0.45\textwidth}{!}{\input{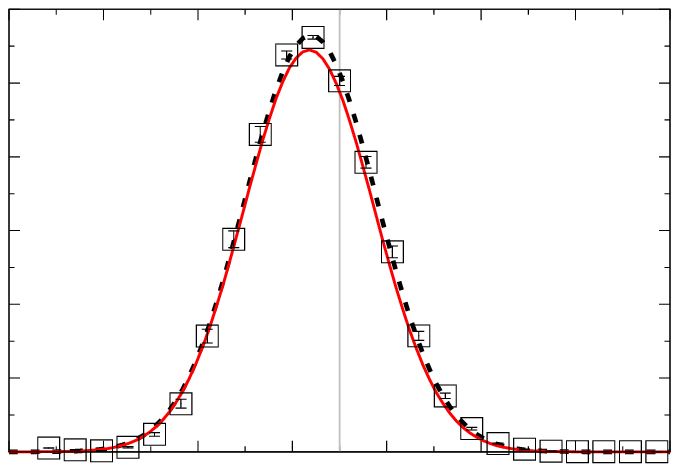}}}
\end{center}
\caption{The observed radial intensity profiles 
(plotted in square boxes with error bars in the $I^*_{\mathrm{obs}}$ direction)
at $\lambda_{\mathrm{obs}}=890\ \mu\mathrm{m}$ of six $PA'$ sectors as examples.
Black dashed lines are Gaussian-fitted curves for the observational data points.
The faint gray line at $r^*=175\ \mathrm{au}$ is drawn 
to emphasize the relative position of $I^*_{0,\mathrm{obs}}$ among the $PA'$ sectors.
Red solid lines are the modeling results using the conventional model for dust opacity. 
The best fit parameters and the disk mass of the sector are written on the top right in each plot. 
The modeling results for $PA'=310^{\circ}-330^{\circ}$,
$PA'=330^{\circ}-350^{\circ}$, and $PA'=350^{\circ}-10^{\circ}$ 
do not meet the criteria in equation (\ref{criteria}).
}
\label{figoribestfit}
\end{figure*}

Figure \ref{figoribestfit} shows the results of the model fitting
for all $PA'$ sectors.
The $15$ radial profiles in $PA'=10^\circ - 310^\circ$
are well reproduced by the models;
their parameters are listed in table \ref{tablebestfitori}.
The mass contained in each $PA'$ sector, $M_\mathrm{sec}$,
is also presented.
However, none of the models succeed in reproducing the 
three radial profiles in $PA'=310^\circ-10^\circ$
(figure \ref{figoribestfit}d, \ref{figoribestfit}e, and \ref{figoribestfit}f).
These three sectors will be described separately in Section \ref{northwest}.
\par
We derived $\Sigma_0=1.12\ \mathrm{g\ cm^{-2}}$
for the brightest $PA'=30^{\circ}-50^{\circ}$,
and $\Sigma_0=8.31\times10^{-3}\ \mathrm{g\ cm^{-2}}$ 
for the faintest $PA'=230^{\circ}-250^{\circ}$.
The contrast in $\Sigma_0$ between these two sectors is about a factor of $130$,
which is twice the value derived by \citet{muto15}.
The apparent discrepancy is discussed in Section \ref{uncertainties}.
Our model might overestimate the temperature in the north ($PA'\approx 11^\circ$) and the south ($PA'\approx 191^\circ$) of the outer disk,
as we did not consider the possible 
shadowing effects caused
by the warped inner disk
\citep{marino15,casassus15}.
As a consequence,
the best fit $\Sigma_0$ for $PA'=350^\circ-10^\circ$ and $PA'=150^\circ-170^\circ$
might be considered as lower limits.
\subsection{The northwest region of the disk}
\label{northwest}
\begin{figure}[htbp]
\input{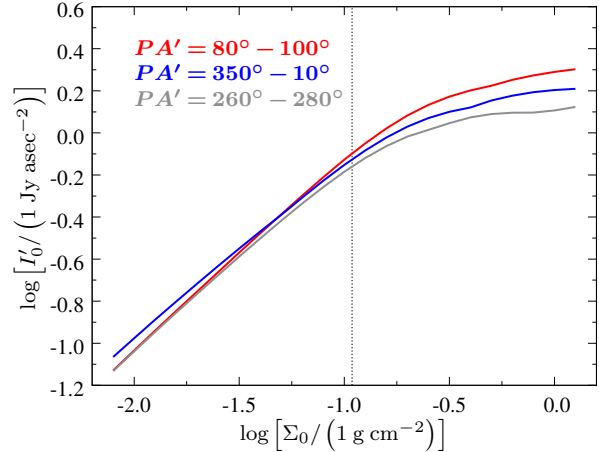}
\centering
\caption{
The peak intensity $I'_0$ achievable in $PA'=80^\circ-100^\circ$ (the far side, red plots), $PA'=350^\circ-10^\circ$ (the major axis, blue plots),
and $PA'=260^\circ-280^\circ$ (the near side, gray plots), regardless of $r'$. 
The $I'_{0}$ values are values before beam convolution.
In the analysis, $w_0$ and $r_0$ are fixed at $27\ \mathrm{au}$ and $175\ \mathrm{au}$, respectively.
The vertical, black dotted line denotes the $\Sigma_0$ value where $\tau_{\mathrm{eff}}=1$. 
}
\label{figivpeakori} 
\end{figure}

We encountered two difficulties in reproducing the three 
radial profiles in $PA'=310^{\circ}-10^{\circ}$.
First, the model intensity is lower than the observed intensity 
even at $\Sigma_0=1.26\ \mathrm{g\ cm^{-2}}$.
In particular, it is $16\%$ lower than $I^*_{0,\mathrm{obs}}$ in $PA'=330^\circ-350^\circ$.
Second, the model profiles in $PA'=310^\circ-350^\circ$
have larger radial extent than those observed even when 
the model width is set to be $w_0=21\ \mathrm{au}$. 
The situation is shown more clearly in figure \ref{figivpeakori},
where the model peak intensity $I'_{0}$ 
is plotted as a function of $\Sigma_0$ for three sectors,
$PA'=80^{\circ}-100^{\circ}$, $350^{\circ}-10^{\circ}$,
and $260^{\circ}-280^{\circ}$;
they correspond to the far side, the major axis, and the near side, respectively.
In $\Sigma_0<0.1~\mathrm{g~cm^{-2}}$,
the peak intensities are roughly proportional to $\Sigma_0$
and are the same for the far and near sides since the disk is optically thin.
This can be confirmed in the model image shown in figure \ref{figdiskmapbcon}a.
In $\Sigma_0 > 0.1~\mathrm{g~cm^{-2}}$,
the slope is shallower for a higher $\Sigma_0$,
and $I'_0$ reaches a ceiling.
The ceiling of $I'_0$
is highest
on the far side, 
followed by the major axis,
and last, on the near side
(figure \ref{figdiskmapbcon}b).
\par
The saturation of the peak intensity can be understood as follows. 
At $\Sigma_0\approx0.1~\mathrm{g~cm^{-2}}$,
the effective optical thickness defined as
\begin{eqnarray}
\label{eqtaueff}
\tau_{\mathrm{eff}} 
             \approx  \Sigma_0 \sqrt{\kappa_{\mathrm{a}}(\kappa_{\mathrm{a}}+\kappa_{\mathrm{s}})}
\end{eqnarray}
is approximately unity,
while the optical thickness by absorption only is
$\tau_\mathrm{a}\approx 0.3$.
The effective optical thickness takes account 
of the diffusion length of a photon
in the presence of scattering (\cite{rybicki79}).
When $\tau_{\mathrm{eff}}$ is less than unity,
the medium is effectively thin or translucent.
Hence, the intensity is roughly proportional
to the column density along the line of sight
when the medium is isothermal
(see Fig. 8 in \cite{muto15} for the temperature distribution).
When $\Sigma_0>0.1\ \mathrm{g\ cm^{-2}}$,
i.e., $\tau_\mathrm{eff}>1$,
the disk is opaque and
the intensity reaches a ceiling.
At the same time,
the peak intensity depends on $PA'$
for a given $\Sigma_0$ model (figure \ref{figdiskmapbcon}b).
The azimuthal dependence comes mainly from the difference
in the temperature of the emitting surface, or the effective photosphere,
which is discussed quantitatively in Section \ref{ipeakdependence}.
\par
When $\Sigma_0 > 1.0~\mathrm{g~cm^{-2}}$,
the intensity profile is also significantly broadened
in the radial direction because the intensity
becomes insensitive to the increase in $\Sigma_0$,
i.e., the profile
becomes
saturated even
before the beam convolution (figure \ref{figdiskmapbcon}).
This situation can also be seen in table \ref{tablebestfitori},
where we have chosen the maximum $\Sigma_0$ so that 
the peak intensity is as high as possible in our parameter space.
The width in the model image,
$w_0^*$, is wider than the observed value, $w^*_\mathrm{obs}$,
by more than $6\%$ in $PA'=310^\circ - 350^\circ$.
Therefore, we conclude that the dust opacity applied here cannot reproduce
the radial intensity profiles in $PA'=310^\circ - 10^\circ$
due to the low intensity ceiling compared to the observation,
as well as its associated effect of 
broadening the intensity profile.
\subsection{Uncertainties in the best fit values}
\label{uncertainties}
\begin{table*}[h!]
\centering
\caption{Fit values for $PA^*=230^{\circ}-250^{\circ}$. 
}
\label{tablepa240error}
\begin{tabular}{ccccccccccc}
\toprule
\multicolumn{5}{c}{\textbf{Model parameters}}   &     \multicolumn{6}{c}{\textbf{Model results}}   \\ 
$\Sigma_0$ & $w_0$ & $r_0$ & $\Sigma_0 w_0$  &   $M_\mathrm{sec}$ &   $I^*_0$ & $w^*_0$ & $r^*_0 $    & $\Delta I^*_0$  &  $\Delta w^*_0$  &  $\Delta r^*_0$   \\ 
$\left( \mathrm{g\ cm^{-2}}\right)$ & $\left( \mathrm{au}\right) $ & $\left( \mathrm{au}\right)$ 
                  &   $\left( \mathrm{g\ cm^{-2}\ au}\right) $ &  $\left( M_{\odot} \right)$ 
                  &   $\left( \mathrm{Jy\ asec^{-2}}\right) $ & $\left( \mathrm{au} \right)$ & $\left( \mathrm{au} \right)$  
                  &   $\left( \times I'_{0,\mathrm{obs}}\right) $ & $\left( \mathrm{au} \right)$ & $\left( \mathrm{au} \right)$ \\
\cmidrule(){1-5} \cmidrule(l){6-11}
$8.31 \times 10^{-3}$                      &  $32$ & $195$ &  $0.27$  & $3.0\times 10^{-6}$ & $0.05$ & $52.3$ & $173.3$  &   $4.6\times10^{-5}$  & $-0.7$ & $0.8$  \\
$8.85\times10^{-3}$ & $29$  & $195$ &  $0.26$  & $2.9\times 10^{-6}$ & $0.05$ & $50.8$ & $173.9$  &   $1.7\times10^{-5}$  & $0.8$ & $0.3$  \\
\bottomrule 
\end{tabular}
\vskip 1cm
\centering
\caption{Fit values for $PA'=30^{\circ}-50^{\circ}$. 
}
\label{tablepa40error}
\begin{tabular}{ccccccccccc}
\toprule
 \multicolumn{5}{c}{\textbf{Model parameters}}   &     \multicolumn{6}{c}{\textbf{Model results}}   \\ 
$\Sigma_0$ & $w_0$ & $r_0$ & $\Sigma_0 w_0$  &   $M_\mathrm{sec}$ &   $I^*_0$ & $w^*_0$ & $r^*_0 $    & $\Delta I^*_0$  &  $\Delta w^*_0$  &  $\Delta r^*_0$   \\ 
$\left( \mathrm{g\ cm^{-2}}\right)$ & $\left( \mathrm{au}\right) $ & $\left( \mathrm{au}\right)$ 
                  &   $\left( \mathrm{g\ cm^{-2}\ au}\right) $ &  $\left( M_{\odot} \right)$ 
                  &   $\left( \mathrm{Jy\ asec^{-2}}\right) $ & $\left( \mathrm{au} \right)$ & $\left( \mathrm{au} \right)$  
                  &   $\left( \times I'_{0,\mathrm{obs}}\right) $ & $\left( \mathrm{au} \right)$ & $\left( \mathrm{au} \right)$ \\
\cmidrule(){1-5} \cmidrule(l){6-11}
$1.12$                      &  $21$ & $173$ &  $23.52$  & $2.6\times 10^{-5}$ & $1.25$ & $50.7$ & $152.6$  &   $2.9\times10^{-5}$  & $0.3$ & $-0.3$  \\
$8.99\times10^{-1}$ & $23$  & $173$ &  $20.68$  & $2.3\times 10^{-5}$ & $1.25$ & $51.7$ & $152.1$  &   $1.8\times10^{-5}$  & $-0.7$ & $0.2$  \\
\bottomrule 
\end{tabular}
\end{table*}
The best fit values cannot be strictly constrained
because of the insufficient angular resolution ($\approx 60\ \mathrm{au}$ in FWHM).
Furthermore,
in the optically thick region
the intensity reaches a ceiling,
resulting in
even larger uncertainties in the results,
especially in $\Sigma_0$.  
The uncertainties in
the optically thin and optically thick sectors are discussed separately,
by using the radial profiles of $PA'=30^{\circ}-50^{\circ}$ 
and $PA'=230^{\circ}-250^{\circ}$ as case studies.
We investigate the
dependence of $M_\mathrm{sec}$ 
on $w_0$,
by treating $\Sigma_0$ in equation (\ref{sigmatimesw})
as the bound variable.
\par
In the regime of optically thin and constant temperature,
\begin{eqnarray}
I^*_0\propto\Sigma_0w_0,
\label{sigmatimesw}
\end{eqnarray}
i.e., the peak intensity is proportional to
the product of $\Sigma_0$ and $w_0$,
or equivalently $M_{\mathrm{sec}}$.
Despite the difference in their $w_0$ values,
the two parameter sets
listed in table \ref{tablepa240error} for
$PA'=230^{\circ}-250^{\circ}$
satisfy the criteria in equation (\ref{criteria}).
The two sets,
each with $w_0=32\ \mathrm{au}$ and $w_0=29\ \mathrm{au}$,
estimate $M_\mathrm{sec}$ to be 
$3.0\times 10^{-6}\ M_\odot$
and
$2.9\times 10^{-6}\ M_\odot$,
respectively.
Both $\Sigma_0$ and $w_0$ 
contain $\pm 10\%$ uncertainties
due to the criteria tolerance in equation (\ref{criteria}),
but $M_\mathrm{sec}$
is uncertain only by approximately $\pm 5\%$.
This is because $I^*_0$ is proportional to $M_\mathrm{sec}$
in these optically thin sectors.
\par
Similarly, 
in table \ref{tablepa40error}
we listed two parameter sets for the optically thick 
$PA'=30^{\circ}-50^{\circ}$.
The difference in terms of 
$M_\mathrm{sec}$ or $\Sigma_0w_0$
is about $11\%$,
which is much larger than that for $PA'=230^\circ-250^\circ$.
The difference in terms of $w_0$ is about $10\%$,
in contrast to the difference in $\Sigma_0$ which reaches $20\%$.
This implies that equation (\ref{sigmatimesw}) breaks down
when the intensity reaches a ceiling.
Moreover, 
if we relax the criteria tolerance 
to follow Muto et al. (2015) (Section \ref{comparecriteria}),
a wider range of $\Sigma_0w_0$ can be obtained as best fit values,
so as the uncertainties in the optically thick sectors
(refer to Appendix 2 in Muto et al. 2015).
This explains the discrepancy between our results
and those derived by Muto et al. (2015):
in this work the $\Sigma_0$ value for $PA'=30^\circ-50^\circ$
is about twice the value derived by Muto et al. (2015),
resulting in the higher $\Sigma_0$ contrast between 
$PA'=30^\circ-50^\circ$ and $PA'=230^\circ-250^\circ$.
The best fit results,
especially the values of $\Sigma_0$ in the optically thick region
in $PA'=10^\circ-130^\circ$ and $PA'=290^\circ-310^\circ$,
contain uncertainties of a factor of $\sim 2$.
Analysis based on observations with a higher angular resolution,
or at a longer wavelength (thus less opaque in the emissions),
should be able to better constrain $\Sigma_0$ and $w_0$.

\section{Discussions}
\label{discussions}
\begin{table*}
\centering
\caption{Derived best fit values using the reduced-scattering model.}
\label{tablebestfitscared}
\captionsetup[subfloat]{position=top}
\begin{tabular}{cccccccc}
\toprule
\multicolumn{1}{c}{\multirow{2}{*}{$PA'$}}        &    $\Sigma_0$  &   $w_{0}$   &   \multicolumn{1}{c}{$r_{0}$}  & $M_\mathrm{sec}$ & $\Delta I^*_{0}$ & $\Delta w^*_0$ & $\Delta r^*_0$ \\
\multicolumn{1}{c}{}        &    $(\mathrm{g\ cm^{-2}})$  &   $(\mathrm{au})$   &   \multicolumn{1}{c}{$(\mathrm{au})$} 
& $(M_{\odot})$ & $\left( \times I^*_{0,\mathrm{obs}}\right)$& $\left(\mathrm{au}\right)$ & $\left(\mathrm{au}\right)$ \\
\cmidrule(){1-1} \cmidrule(l){2-5} \cmidrule(l){6-8}
$10^{\circ}-30^{\circ}$  &  $2.76 \times 10^{-1}$  &  $31$  &  $173$  &  $1.0 \times 10^{-4}$ & $-4.0\times10^{-5}$ & $+0.9$ & $-0.0$\\   
$ 30^{\circ}-50^{\circ}$  &  $3.34 \times 10^{-1}$  &  $31$  &  $173$  &  $1.1 \times 10^{-4}$ & $-2.6\times10^{-5}$ & $-0.6$ & $-0.9$\\   
$ 50^{\circ}-70^{\circ}$  &  $3.33 \times 10^{-1}$  &  $29$  &  $173$  &  $9.8 \times 10^{-5}$ & $-2.7\times10^{-5}$ & $-0.8$ & $-0.5$\\  
$ 70^{\circ}-90^{\circ}$  &  $2.46 \times 10^{-1}$  &  $27$  &  $169$  &  $6.1 \times 10^{-5}$ & $-2.2\times10^{-5}$ & $-0.5$ & $-0.6$\\   
$\phantom{1}90^{\circ}-110^{\circ}$  &  $1.84 \times 10^{-1}$  &  $23$  &  $167$  &  $3.9 \times 10^{-5}$ & $-3.1\times10^{-5}$ & $+0.0$ & $-0.9$\\   
$110^{\circ}-130^{\circ}$  &  $1.24 \times 10^{-1}$  &  $23$  &  $165$  &  $2.8 \times 10^{-5}$ & $-2.8\times10^{-5}$ & $+0.1$ & $+0.2$\\   
$130^{\circ}-150^{\circ}$  &  $6.94 \times 10^{-2}$  &  $27$  &  $167$  &  $2.0 \times 10^{-5}$ & $-2.7\times10^{-5}$ & $-0.7$ & $-0.7$\\   
$150^{\circ}-170^{\circ}$  &  $3.32 \times 10^{-2}$  &  $29$  &  $171$  &  $1.1 \times 10^{-5}$ & $-1.9\times10^{-5}$ & $-0.2$ & $-0.9$\\  
$170^{\circ}-190^{\circ}$  &  $1.51 \times 10^{-2}$  &  $31$  &  $177$  &  $5.7 \times 10^{-6}$ & $-1.0\times10^{-5}$ & $-0.4$ & $-0.5$\\   
$190^{\circ}-210^{\circ}$  &  $1.07 \times 10^{-2}$  &  $29$  &  $187$  &  $3.9 \times 10^{-6}$ & $-9.1\times10^{-6}$ & $+0.7$ & $+0.6$\\   
$210^{\circ}-230^{\circ}$  &  $8.97 \times 10^{-3}$  &  $32$  &  $193$  &  $3.5 \times 10^{-6}$ & $-3.5\times10^{-5}$ & $-0.5$ & $-0.2$\\   
$230^{\circ}-250^{\circ}$  &  $8.28 \times 10^{-3}$  &  $32$  &  $195$  &  $3.0 \times 10^{-6}$ & $-1.5\times10^{-5}$ & $-0.6$ & $+0.1$\\   
$250^{\circ}-270^{\circ}$  &  $2.80 \times 10^{-2}$  &  $23$  &  $185$  &  $6.5 \times 10^{-6}$ & $-4.4\times10^{-5}$ & $+0.1$ & $-0.3$\\   
$270^{\circ}-290^{\circ}$  &  $7.26 \times 10^{-2}$  &  $27$  &  $183$  &  $2.0 \times 10^{-5}$ & $-7.5\times10^{-7}$ & $-0.3$ & $+0.9$\\   
$290^{\circ}-310^{\circ}$  &  $1.46 \times 10^{-1}$  &  $27$  &  $181$  &  $4.2 \times 10^{-5}$ & $-8.1\times10^{-7}$ & $-0.1$ & $-0.6$\\   
$310^{\circ}-330^{\circ}$  &  $2.37 \times 10^{-1}$  &  $29$  &  $177$  &  $7.7 \times 10^{-5}$ & $-1.1\times10^{-5}$ & $+0.0$ & $+0.3$\\   
$330^{\circ}-350^{\circ}$  &  $2.91 \times 10^{-1}$  &  $29$  &  $175$  &  $1.0 \times 10^{-4}$ & $-2.2\times10^{-5}$ & $+0.7$ & $-0.4$\\
$350^{\circ}-10^{\circ}\phantom{3}$    &  $2.59 \times 10^{-1}$  &  $31$  &  $173$  &  $9.6 \times 10^{-5}$ & $-2.2\times10^{-5}$ & $+0.6$ & $-0.9$\\   
\bottomrule
\end{tabular}
\end{table*}
\begin{figure*}
\begin{center}
\subfloat[]{\resizebox{0.45\textwidth}{!}{\input{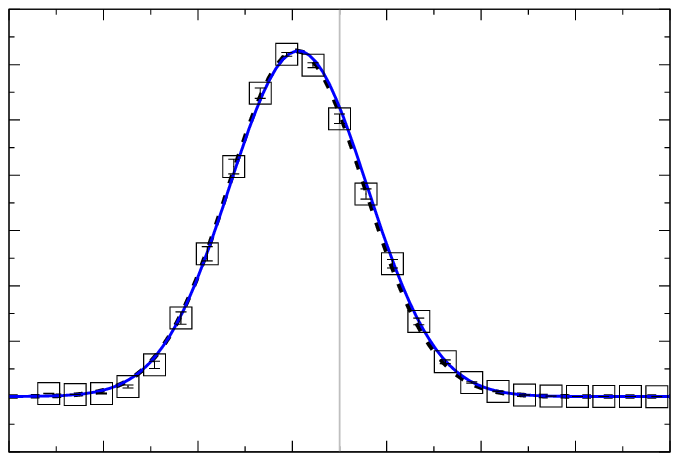}}}
\hspace{10mm}
\subfloat[]{\resizebox{0.45\textwidth}{!}{\input{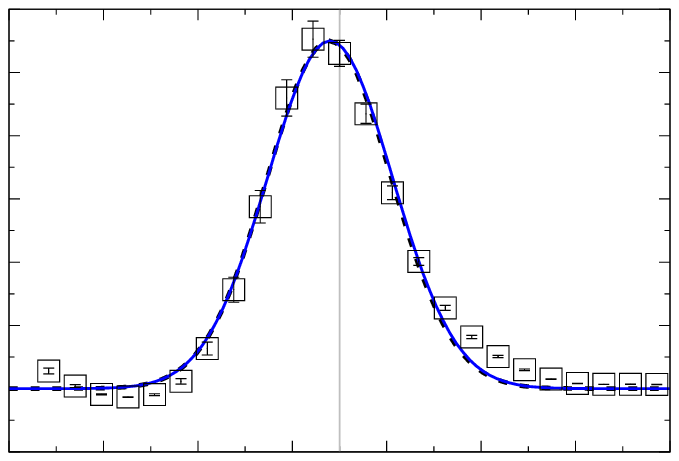}}}
\\
\subfloat[]{\resizebox{0.45\textwidth}{!}{\input{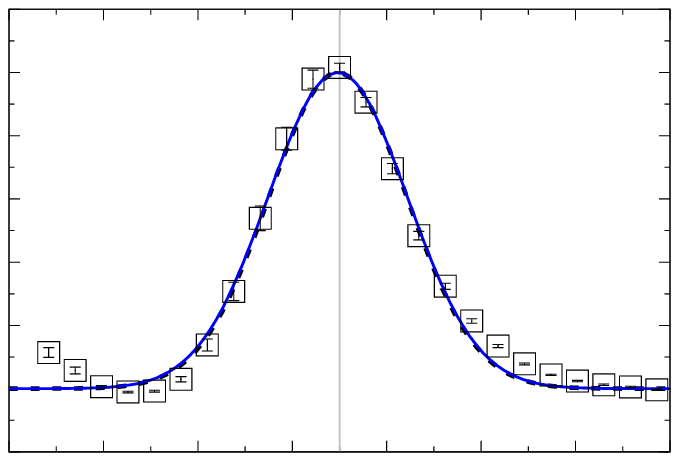}}}
\hspace{10mm}
\subfloat[]{\resizebox{0.45\textwidth}{!}{\input{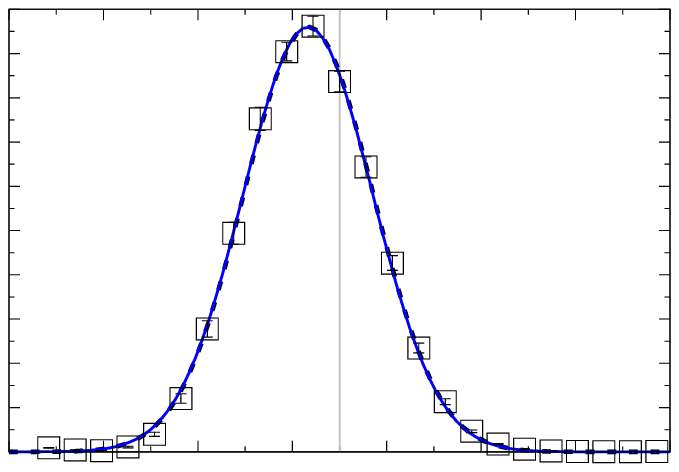}}}
\\
\subfloat[]{\resizebox{0.45\textwidth}{!}{\input{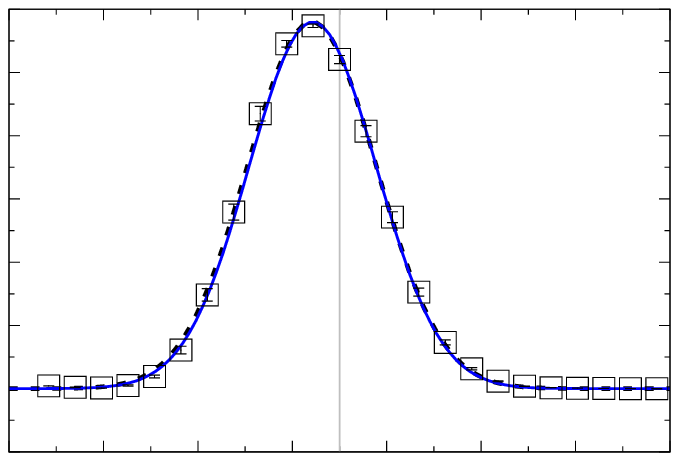}}}
\hspace{10mm}
\subfloat[]{\resizebox{0.45\textwidth}{!}{\input{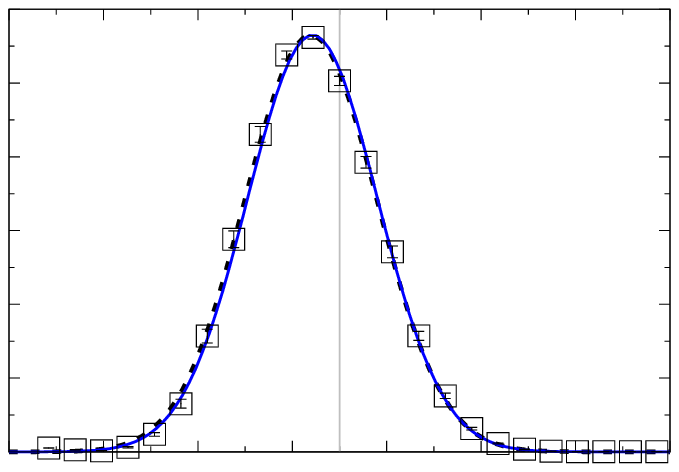}}}
\end{center}
\caption{Similar to figure \ref{figoribestfit}, but the results (blue solid lines) here are derived by the reduced-scattering model.
The radial profiles of $PA'=310^{\circ}-330^{\circ}$, $PA'=330^{\circ}-350^{\circ}$, and $PA'=350^{\circ}-10^{\circ}$
are well reproduced with the reduced-scattering model.}
\label{figscabestfit}
\end{figure*}
\begin{figure}[h!]
\input{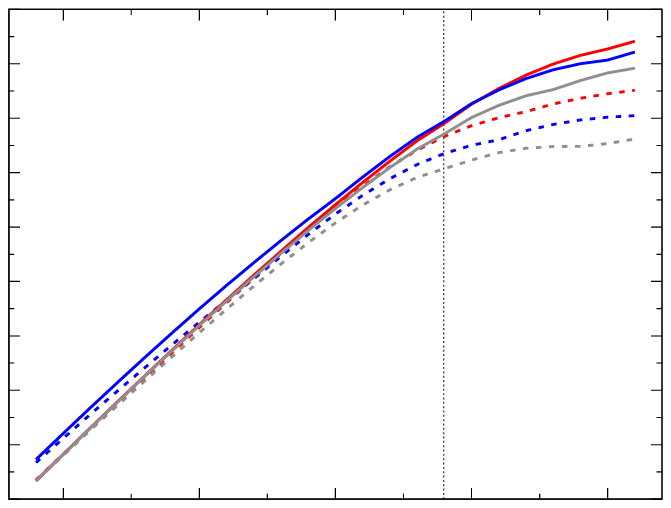}
\centering
\caption{
The maximum surface brightness $I'_0$ achievable in $PA'=80^\circ-100^\circ$ (red plots), $PA'=350^\circ-10^\circ$ (blue plots),
and $PA'=260^\circ-280^\circ$ (gray plots), regardless of $r'$. 
Solid and dashed lines are the results using the reduced-scattering and conventional models, respectively. 
The $I'_{0}$ values are values before beam convolution.
The vertical, black dotted line denotes the $\Sigma_0$ value where $\tau_{\mathrm{eff}}=1$ in the reduced-scattering model.
}
\label{figivpeak} 
\end{figure}
\subsection{Reducing scattering opacities}
\label{scaredmodel}
\begin{figure*}[h!]
\begin{center}
\input{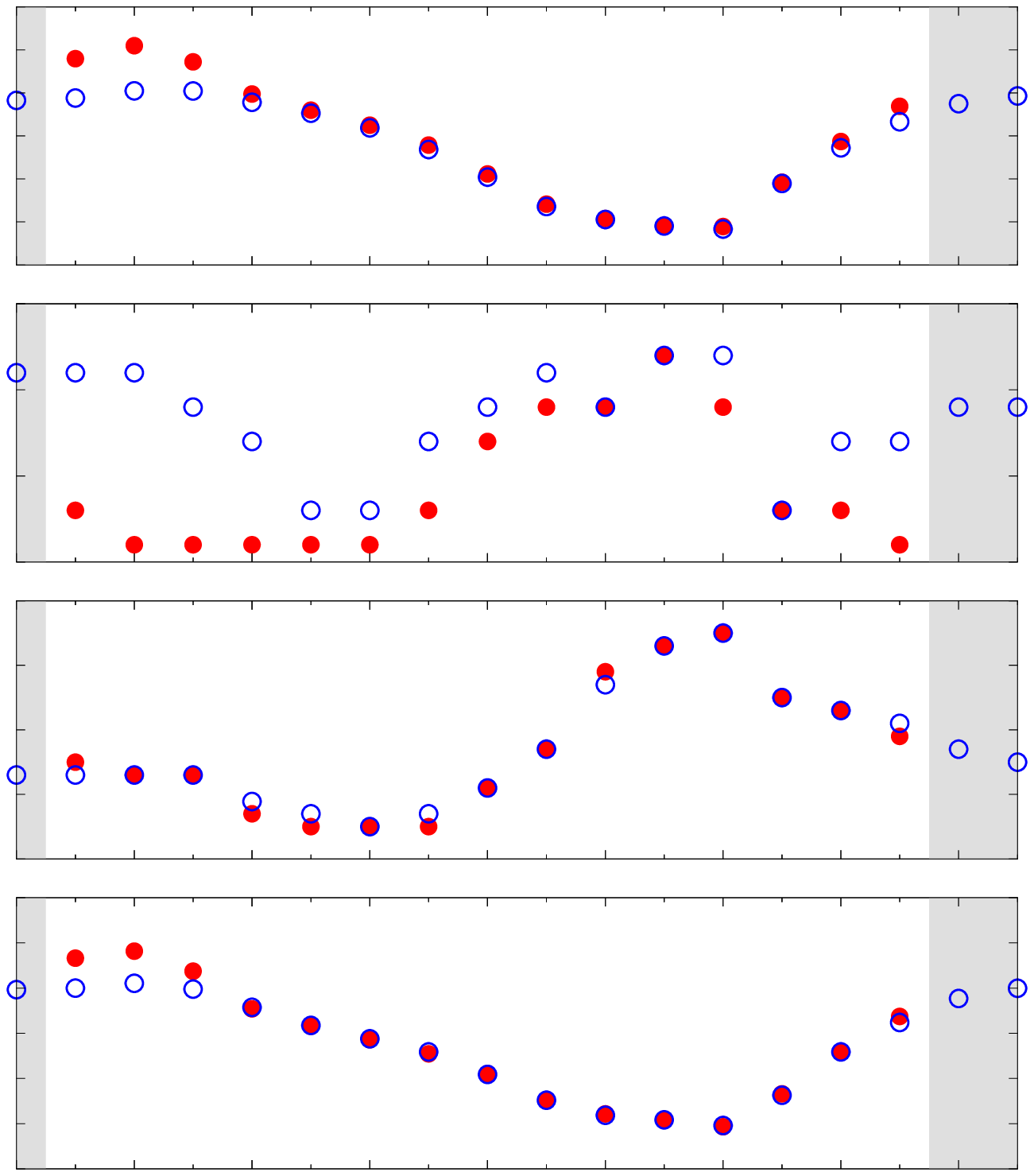}
\end{center}
\caption{
The best fit parameters, $\Sigma_0$, $w_0$, $r_0$, and $M_\mathrm{sec}$, distributions along the $PA'$ direction,
derived using the conventional (red filled circles) and the reduced-scattering model (blue open circles).
The gray shaded portions indicate the $PA'$ sectors where the radial profiles cannot be reproduced by 
the conventional model.
}
\label{figpasum}
\end{figure*}
Figure \ref{figivpeakori} shows that $I'_0$ depends on both $\Sigma_0$ and $PA'$,
and 
it reaches a ceiling when $\tau_{\mathrm{eff}}$ exceeds unity.
The low intensity ceiling
is a consequence of the large scattering optical depth.
Although the conventional properties of the dust grains
are applied in the model,
their actual opacity can be highly uncertain;
they might be aggregates composed of monomers and have different composition \citep{tazaki16},
so that either the scattering cross section might have been overestimated
or the asymmetry parameter been underestimated, or both.
Therefore, as a tentative approach to reproduce 
the radial profiles in the northwestern region, $PA'=310^{\circ}-10^{\circ}$,
we reduce the scattering opacity 
across all the wavelengths by $90\%$, as depicted in figure \ref{figopacity}.
The scattering opacity at $\lambda_{\mathrm{obs}}$ is then reduced to $2.62\ \mathrm{cm^{2}\ g^{-1}}$,
which approximates the absorption opacity.
The purpose is
to reduce the scattering optical depth,
thus elevating the ceiling of $I'_0$.
The absorption opacity of the dust particles is kept the same.
Therefore the albedo $\eta$ is thus effectively reduced.
In the following, we denote this model by \textquotedblleft reduced-scattering model\textquotedblright, 
in contrast to the  \textquotedblleft conventional model\textquotedblright\ which uses the conventional scattering opacity.
We apply the reduced-scattering model to reproduce not only
the radial profiles of the northwestern region,
but also to all $PA'$ sectors in order to study the relevant effects.
\par
The results are shown in figure \ref{figscabestfit} and in table \ref{tablebestfitscared}.
We successfully reproduce the radial profiles in 
all sectors,
including those in $PA'=310^{\circ}-10^{\circ}$
(figure \ref{figscabestfit}d, \ref{figscabestfit}e, and \ref{figscabestfit}f).
In addition,
the contrast in $\Sigma_0$ between 
$PA'=30^\circ-50^\circ$ and
$PA'=230^\circ-250^\circ$ is reduced
to about a factor of $40$,
which is closer to the contrast in observed intensity.
Using the reduced-scattering model,
we plot $I'_0$
as a function of $\Sigma_0$
in figure \ref{figivpeak}. 
The results are essentially similar to those of the conventional model,
but 
$I'_0$ reaches a ceiling
only at $\Sigma_0 \approx 0.25\ \mathrm{g\ cm^{-2}}$,
at which $\tau_{\mathrm{eff}}\approx1$.
Besides, the reduced-scattering model is brighter 
than the conventional model in all the cases,
especially when $\Sigma_0$ exceeds $0.1\ \mathrm{g\ cm^{-2}}$,
where $\tau_{\mathrm{eff}}$ is approximately $1$ in the conventional model
but only $0.3$ in the reduced-scattering model.
\par
We summarize the best fit parameters and
the dust masses derived from the conventional and the reduced-scattering models in figure \ref{figpasum}.
Compared to the conventional model,
the reduced-scattering model estimates smaller $\Sigma_0$
but larger $w_0$ in the northern region.
Furthermore, the $\Sigma_0$ distribution
in $PA'=310^\circ-70^\circ$
is about $0.3\ \mathrm{g\ cm^{-2}}$ on average,
contrary to the conventional model which derives
about $0.9\ \mathrm{g\ cm^{-2}}$ for sectors
in $PA'=10^\circ - 70^\circ$.
When scattering is dominant, the disk appearance is faint and indistinct,
restricting only 
the model of small $w_0$ and large $\Sigma_0$ to reproduce $w^*_{0,\mathrm{obs}}$.
Reducing the scattering opacities sharpens the disk outline and thus allows us to 
reproduce the same $I^*_0$ with a larger $w_0$ but at a smaller $\Sigma_0$, 
as predicted by equation (\ref{sigmatimesw}).
Since the reduced-scattering model
is yet to be optically thick for 
the intensity to reach a ceiling,
it estimates about half the mass derived from the conventional model.
The sectors within $PA'=130^\circ - 290^\circ$
show no
significant changes in
their best fit parameters like
those in the northern region;
their $\tau_{\mathrm{eff}}$ values are below unity
so that the intensity is not much different between the conventional and reduced-scattering models.
As a conclusion, reducing scattering opacity alters the results only in the optically thick northern sectors.
\subsection{Intensity Dependence on Scattering Opacity and $PA'$}
\label{ipeakdependence}
\begin{figure*}
\begin{center}
\subfloat[$\theta=-27^{\circ}$]{
\resizebox{0.36\textwidth}{!}{\input{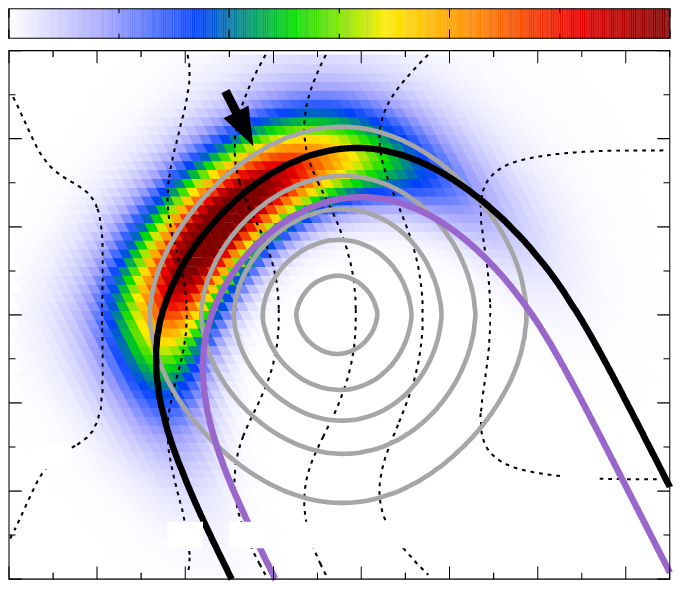}}
\hspace{10mm}
\resizebox{0.36\textwidth}{!}{\input{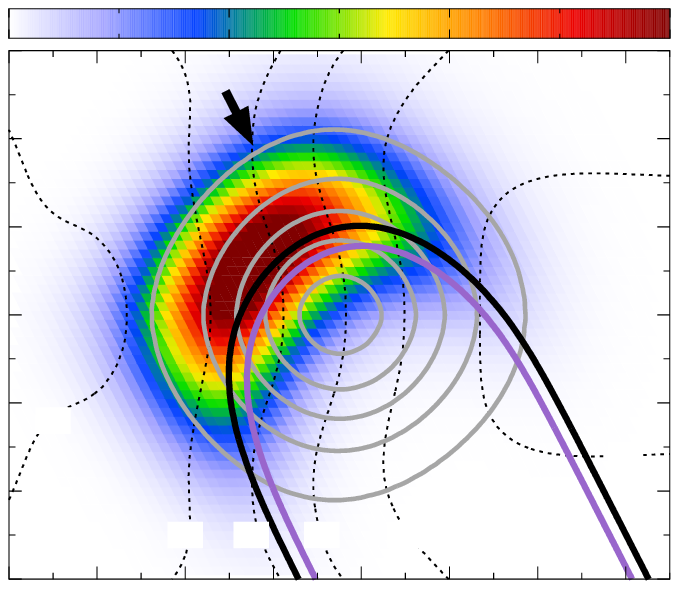}}
}
\\
\subfloat[$\theta=0^{\circ}$]{
\resizebox{0.36\textwidth}{!}{\input{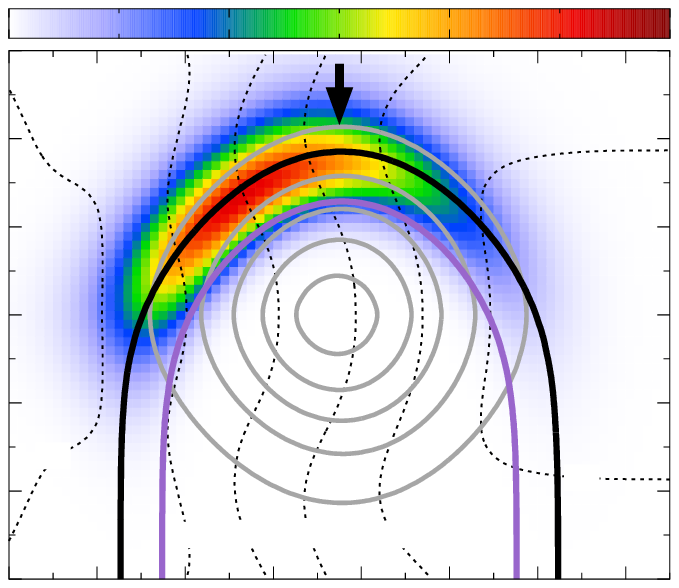}}
\hspace{10mm}
\resizebox{0.36\textwidth}{!}{\input{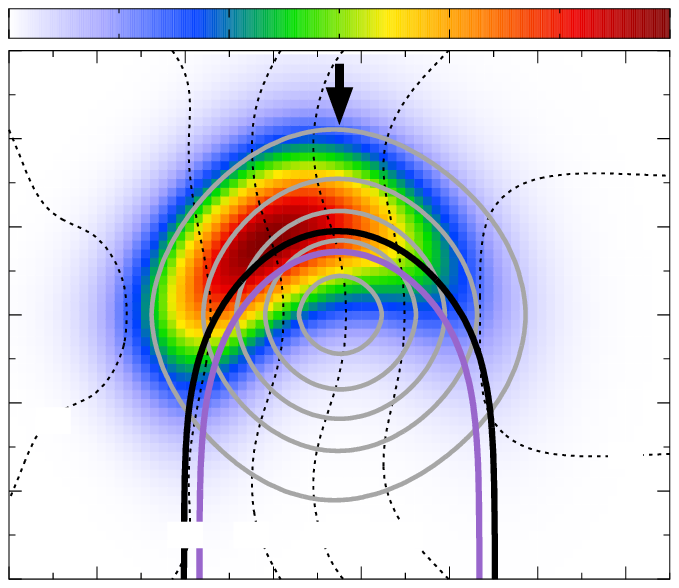}}
}
\\
\subfloat[$\theta=+27^{\circ}$]{
\resizebox{0.36\textwidth}{!}{\input{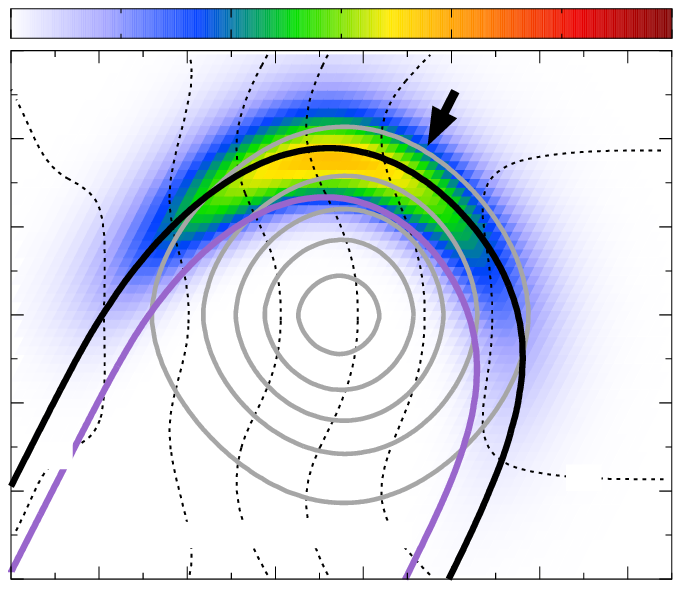}}
\hspace{10mm}
\resizebox{0.36\textwidth}{!}{\input{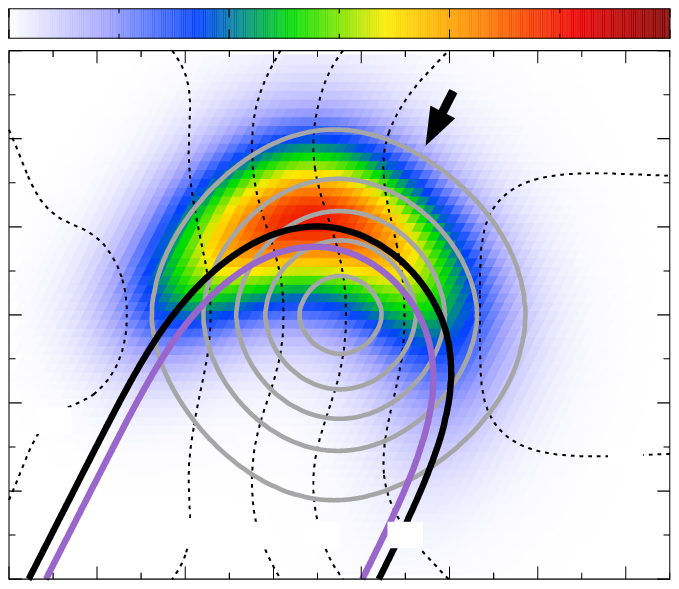}}
}
\end{center}
\caption{
The $(\kappa_\mathrm{a}+\kappa_\mathrm{s}) \rho S_{\nu} \exp(-\tau')$ distributions of the conventional (left panels) 
and the reduced-scattering model (right panels), at a common $\Sigma_0=1.00\ \mathrm{g\ cm^{-2}}$.
The color scale is common for each panel.
The top, middle, and bottom panels are the distributions at viewing angle 
$\theta=-27^{\circ}$, $0^{\circ}$, and $+27^{\circ}$, respectively, 
as indicated by the black arrows.
The black dotted contours are the temperature (units in $\mathrm{K}$).
The gray solid contours are the densities drawn at
$\rho(r,z)=0.1\ ,\ 0.3\ ,\ 0.5 \ ,\ 0.7,\ 0.9$ of
$\rho_{0}(175\ \mathrm{au},0\ \mathrm{au})=1.3\times 10^{-15}\ \mathrm{g\ cm^{-3}}$,
the peak density in the model.
The surface where $\tau' = 1$ and $\tau_\mathrm{eff} = 1$
are indicated by the thick black and the purple curves, respectively.
}
\label{figsvtaunm22}
\end{figure*}
In this subsection we discuss
why the reduced-scattering model can reproduce
the ALMA observation while 
the conventional model cannot.
We also discuss the origin of the azimuthal dependence
of the peak intensity.
For these purposes, 
we evaluate the model intensity quantitatively.
\par
The solution of equation (\ref{equationraytracing}) is expressed as
\begin{eqnarray}
I' (r;PA')= (\kappa_\mathrm{a}+\kappa_\mathrm{s}) \int^{+\infty}_{-\infty} \rho(r,z) S_{\nu}(r,z) e^{-\tau'} dz',
\end{eqnarray}
where $\tau'$ denotes the optical depth measured from the observer 
to the point $(r,z)$ in the disk 
along the line of sight $z'$, defined by
\begin{eqnarray}
\tau' (r,z;PA') = (\kappa_\mathrm{a}+\kappa_\mathrm{s}) \int^{z}_{{+\infty}} \rho dz'.
\end{eqnarray}
The observer is located at $z=+\infty$.
The observable photon density, 
$(\kappa_\mathrm{a}+\kappa_\mathrm{s}) \rho S_{\nu} \exp(-\tau')$,
denotes the emissivity per unit length 
in which absorption and scattering are taken into account
by the attenuation factor $\exp(-\tau')$.
Figure \ref{figsvtaunm22} shows the observable photon density 
in unit of $\mathrm{Jy\ asec^{-2}\ au^{-1}}$
for the model of $\Sigma_0=1.00\ \mathrm{g\ cm^{-2}}$,
$w_0=27\ \mathrm{au}$, and $r_0=175\ \mathrm{au}$.
The left and right panels are based on the conventional 
and the reduced-scattering models, respectively.
From top to bottom,
the line of sight is offset from the $z$-axis by 
$\theta=-27^\circ$, $0^\circ$, and $+27^\circ$.
Since the inclination of the disk is 
$i=27^\circ$, 
the top panels correspond to $PA'=90^\circ$ (the far side)
while the bottom panels to $PA'=270^\circ$ (the near side).
The middle panels where $\theta=0^\circ$ 
shows the observable photon density of a face-on disk;
they do not actually correspond to $PA'=0^\circ$
because
our line of sight does not lie in the $(r,z)$ plane.
However, the results
can be regarded as a good substitute
for the $PA'=0^\circ$ case since the inclination angle is small.
\par
In each panel of figure \ref{figsvtaunm22},
the gray solid and black dotted contours denote
the density and temperature, respectively.
The density and temperature are almost unchanged
by the reduction of scattering.
The observable photon density,
on the other hand,
is greatly different 
between the conventional and the scattering-reduced models,
and it also varies with $\theta$.
The thick black and purple curves denote
the location of $\tau'=1$ and $\tau_\mathrm{eff}=1$, respectively;
the location of $\tau_\mathrm{eff}=1$ is plotted
by scaling $\tau'$ by a factor of $(1-\eta)^{-1/2}$.
\par
The observed photons come mainly
from the region of $\tau'=1$,
where $\tau_\mathrm{eff}$ is still smaller than unity.
This is because
while most of the photons are often emitted from the
$\tau_\mathrm{eff}=1$ surface,
the observed photons are scattered around
$\tau'=1$ surface before reaching the observer.
In the conventional model,
the difference between $\tau_\mathrm{eff}=1$ and $\tau'=1$ is
large,
and its observable photon emitting area is confined in a narrow region
near $\tau'=1$ (left panels in figure \ref{figsvtaunm22}).
When the reduced-scattering model is applied,
the regions of $\tau_\mathrm{eff}=1$ and $\tau'=1$
become closer
and the observable photon density extends over a wider area
(right panels in figure \ref{figsvtaunm22}).
The narrowing of the emitting area in the conventional model
is equivalent to the decrease in propagation speed of the photons 
caused by the dominant scattering.
Considering that the source function of radiation is similar in both models,
the intensity is higher in the reduced-scattering model
because the line of sight
crosses a longer path in the area of high observable photon density. 
It should be noted that the model peak intensity is lower 
than that of the blackbody radiation for the dust temperature
in the presence of scattering.
\par
The effects of scattering on the intensity
is serious only when 
$\kappa_\mathrm{s} > \kappa_\mathrm{a}$.
Further reduction in the scattering opacity
will increase the intensity only by a little,
thus we cannot set a lower limit on the scattering opacity from our modeling.
Nevertheless, the scattering opacity of the dust grains should be appreciable since
the pattern of polarization in the disk at $\lambda = 890\ \mu\mathrm{m}$,
recently revealed by ALMA, seems consistent with the case of self-scattering
\citep{kataoka15,kataoka16}. 
\par
The peak intensity shows azimuthal dependence when the disk is inclined,
and the observed intensity strongly depends on 
the temperature of the effective photosphere
when $\tau'$ (or $\tau_\mathrm{eff}$) exceeds unity (figure \ref{figdiskmapbcon}b).
The observable photon density at a given point depends on $\theta$
through the attenuation factor $\exp(-\tau')$,
and the peak density decreases in the order of which $\theta=-27^\circ$,
$0^\circ$, and $+27^\circ$.
Since the temperature gradient is due to the heating from the central star,
the decrease in the observable photon density 
directly corresponds to the decrease in the temperature of the effective photosphere
that we can observe.
This is the origin of the azimuthal dependence shown in 
figures \ref{figdiskmapbcon}b and \ref{figivpeakori} when $\tau_\mathrm{eff} > 1$,
in which the far side is the highest in intensity while the near side is the lowest.
\par
As mentioned at the beginning of Section \ref{modeling},
our 2D axisymmetric model 
might not be able to describe the scattered light properly
in the crescent disk,
especially when scattering is dominant. 
We discuss the validity of 
the axisymmetric disk by using figure \ref{figsvtaunm22}.
In the figure,
the difference between
$\tau'=1$ and $\tau_\mathrm{eff}=1$
corresponds to the typical length travelled by a photon before
escaping from the disk. 
Even when the albedo is high as in the conventional model,
this length is approximately $10~\mathrm{au}$,
much shorter than the arc length subtended by the $20^\circ$-sector
at $r^*_{0,\mathrm{obs}} \approx 160~\mathrm{au}$，
which is approximately $56~\mathrm{au}$.
This implies that the contribution
of scattered radiation from other sectors is negligible,
and the 2D axisymmetric model
is a good approximation to reproduce
the asymmetry intensity distribution of the disk.

\section{Summary}
\label{summary}
We derived the dust distribution in the crescent disk around HD 142527
by reproducing the observed radial intensity profiles of the continuum emission
at $890\ \mu\mathrm{m}$ obtained by ALMA Cycle 0.
The radial distribution of surface density is assumed  
to be a Gaussian function
whose peak
and radial width are denoted as
$\Sigma_0$ and $w_0$, respectively.
The radiative and hydrostatic equilibria in the disk are solved by radiative transfer calculations.
We summarize our results as follows:
\begin{enumerate}
\item
	We first adopt the 
	absorption opacity $\kappa_{\mathrm{a}}$ 
	and scattering opacity $\kappa_{\mathrm{s}}$
	as those of the compact spheres 
	having composition consistent with the solar elemental abundance
	and a power-law size distributionß
	with the maximum radius of $1\ \mathrm{mm}$.
	At the modeling wavelength $890\ \mu\mathrm{m}$,
	$\kappa_{\mathrm{s}}$ is about $10$ times larger than $\kappa_{\mathrm{a}}$.	
	Using these conventional opacities,ß
	we cannot reproduce the radial intensity profiles
	in the northwestern region of the disk,
	observation of which shows high intensity and is located in the near side of the disk.
	The model intensity in this region
	reaches a lower ceiling 
	than the observed peak intensity
	due to heavy scattering.
\item
	When the scattering opacity is reduced
	to $10\%$ from the conventional value,
	the observed intensity distributions
	are reproduced successfully 
	in all the position angles, including those in the northwestern region.
	The best fit values depend little on the scattering opacity
	in the southern region where the disk is optically thin.
	On the other hand, $\Sigma_0$ 
	is derived to be $\sim 50\%$ smaller,
	while $w_0$ becomes $\sim50\%$ wider
	in the optically thick northern region.
	The contrast in $\Sigma_0$ between
	$PA=11^\circ-31^\circ$ and $PA=211^\circ-231^\circ$,
	the brightest and the faintest regions in the dust continuum emission,
	is derived to be $\approx 40$, much smaller than the value 
	for the cases of conventional opacities ($70-130$).
\item 
	Detailed inspection of our model shows that
	in the case of inclined, optically thick disk,
	the temperature on the emitting surface
	depends on the position angle even when the disk is axisymmetric. 
	As a result, 
	the emergent peak intensity also varies azimuthally.
	By reducing the scattering opacity,
	we derived smaller $\Sigma_0$ to reproduce
	the observed intensity in the northern region.
	This is not only because the location of emitting surface
	gets deeper into the disk,
	but also the portion of the disk that contributes to the observed intensity becomes larger.
\end{enumerate}

\section{Funding}
This work is supported by MEXT KAKENHI Nos. 23103004,
24103504, and 26103702.

\section{Acknowledgments}
We thank Hideko Nomura for providing us the 
dust opacity used in this paper.
We also thank Satoshi Okuzumi and Akimasa Kataoka for their valuable comments.
This paper makes use of the following ALMA data: ADS/JAO.ALMA\#2011.0.00318.S.
ALMA is a partnership of ESO (representing its member states),
NSF (USA) and NINS (Japan), together with NRC (Canada), NSC and ASIAA (Taiwan),
and KASI (Republic of Korea), in cooperation with the Republic of Chile.
The Joint ALMA Observatory is operated by ESO, AUI/NRAO and NAOJ.
A part of data analysis was carried out on common use data analysis computer system at the Astronomy Data Center of NAOJ.
K.L. Soon is supported by Japanese Government Scholarship.

\appendix
\section{The M1 model}
\label{appendix1}
Here, we summarize the M1 model,
the approximation method
used to solve the radiation transfer
in our disk model \citep{gonzalez07}.
At coordinate $\boldsymbol{x}$ and at time $t$,
the radiation transfer equation for the specific intensity $I_\nu(\boldsymbol{x},t;\boldsymbol{n})$
along direction $\boldsymbol{n}$
is
\begin{eqnarray}
\label{eqradiationtransfer}
&{}&\frac{1}{c} \frac{\partial I_\nu (\boldsymbol{x},t;\boldsymbol{n})}{\partial t} + \boldsymbol{n} \cdot \nabla  I_\nu (\boldsymbol{x},t;\boldsymbol{n}) = \nonumber\\
&{}&\quad\quad\quad \kappa_\mathrm{a} \rho(\boldsymbol{x},t) B_\nu(\boldsymbol{x},t) 
	-(\kappa_\mathrm{a}+\kappa_\mathrm{s}) \rho(\boldsymbol{x},t)I_\nu (\boldsymbol{x},t;\boldsymbol{n}) \nonumber\\
&{}& \quad\quad\quad\quad\quad + \kappa_\mathrm{s} \rho(\boldsymbol{x},t) \int_{4\pi} I_\nu (\boldsymbol{x},t;\boldsymbol{n'}) d\boldsymbol{n'},
\end{eqnarray}
where $\kappa_\mathrm{a}$ and $\kappa_\mathrm{s}$
denote the absorption and scattering opacities at frequency $\nu$, respectively,
while $c$ and $\rho$ the speed of light and the matter density.
On the right hand side of the equation,
the first term
states that the emissivity of matter is proportional to the blackbody radiation
$B_\nu$ at temperature $T$,
while the second term is read as the extinction by the matter.
The last term, on the other hand,
represents the intensity scattered from $\boldsymbol{n'}$ into $\boldsymbol{n}$.
In the M1 model,
anisotropic scattering can be accounted for
by incorporating the factor $(1-g)$ in the scattering cross section $C_\mathrm{s}$
(see Section \ref{dustmodel}).
Integrating equation (\ref{eqradiationtransfer}) and 
the dot product of equation (\ref{eqradiationtransfer}) between $\boldsymbol{n}$ over solid angle yields
\begin{eqnarray}
\frac{\partial E_\nu}{\partial t} &+& \nabla \cdot \boldsymbol{F}_\nu = \kappa_\mathrm{a} \rho \left( 4\pi B_\nu - cE_\nu \right), \label{eqenu} \\
\frac{\partial \boldsymbol{F}_\nu}{\partial t} &+& c^2 \nabla \cdot \mathcal{P}_\nu 
= - (\kappa_\mathrm{a} + \kappa_\mathrm{s}) c \rho \boldsymbol{F}_\nu, \label{eqfnu}
\end{eqnarray}
respectively,
where, $E_\nu$, $\boldsymbol{F}_\nu$, and $\mathcal{P}_\nu$ are defined as follows:
\begin{eqnarray}
E_\nu &=& \int_{4\pi} I_\nu (\boldsymbol{x},t;\boldsymbol{n}) d\boldsymbol{n}, \nonumber\\
\boldsymbol{F}_\nu &=& \int_{4\pi} \boldsymbol{n} I_\nu (\boldsymbol{x},t;\boldsymbol{n}) d\boldsymbol{n}, \nonumber\\
\mathcal{P}_\nu &=& \int_{4\pi} \boldsymbol{n} \boldsymbol{n} I_\nu (\boldsymbol{x},t;\boldsymbol{n}) d\boldsymbol{n}, \nonumber
\end{eqnarray}
i.e., they are the zeroth, the first, and the second angular moments of the intensity, respectively.
In writing the conservation equations (\ref{eqenu}) and (\ref{eqfnu}),
we assume coherent scattering whereby the last term on
the right hand side of equation (\ref{eqradiationtransfer})
vanishes.
The two conservation equations are closed
by relating $\mathcal{P}_\nu$ and $E_\nu$ via
\begin{eqnarray}
\mathcal{P}_{\nu} =
\left(   \frac{1 - \chi_\nu}{2} \mathcal{I}  + \frac{3 \chi_\nu - 1 }{2}  
   \frac{\boldsymbol{f}_\nu\boldsymbol{f}_\nu}{|\boldsymbol{f}_\nu |^2}  \right) E_\nu,
\end{eqnarray}
where 
\begin{eqnarray}
\chi_\nu  &=&  \frac{3 + 4  |\boldsymbol{f} _\nu|^2}{5+\sqrt{4 - 3 |\boldsymbol{f} _\nu |^2}}, \nonumber \\
\boldsymbol{f}_\nu   &\equiv& \frac{\boldsymbol{F}_\nu}{cE_\nu}, \nonumber
\end{eqnarray}
and $\mathcal{I}$ the identity matrix.
\par
As described in Section \ref{radiationtransfer},
we use the cylindrical coordinates $(r,\varphi,z)$
in which the disk midplane coincides with $z=0$
and the star locates at the origin.
The disk is symmetric around $z$-axis
and with respect to the midplane. 
We consider the star as the heat source for the disk,
and use $226$ colors within $0.1 \mathrm{\ \mu m} \leq \lambda \leq 3.16 \mathrm{\ mm}$
(equivalent spectral resolution of $\Delta \log\lambda=0.02$)
in the calculation. 
Decomposing the radiation energy density and
the flux into the components each from the star and the disk
\citep{kanno13},
we write
\begin{eqnarray}
E_\nu(r,z)&=&E_{\nu,\mathrm{star}}(r,z)+E_{\nu,\mathrm{disk}}(r,z),\\
\boldsymbol{F}_\nu(r,z)&=&\boldsymbol{F}_{\nu,\mathrm{star}}(r,z)+\boldsymbol{F}_{\nu,\mathrm{disk}}(r,z).
\end{eqnarray}
The star is assumed to be a blackbody of effective temperature $T_\mathrm{eff}$,
and its radiation energy density and flux intercepted by the disk at $(r,z)$ are evaluated as
\begin{eqnarray}
&{}& E_{\nu,\mathrm{star}} (r,z)  = \frac{\pi R_\mathrm{star}^2}{c \left( r^2 + z^2\right)} B_\nu \left( T_\mathrm{eff} \right) \exp \left[ -\tau (r,z) \right], \\
&{}& \boldsymbol{F}_{\nu,\mathrm{star}} (r,z) = \frac{c}{\sqrt{r^2 + z^2}} \left( \begin{array}{c} r \\ z \end{array} \right) E_{\nu,\mathrm{star}} (r,z),
\end{eqnarray}
where $\tau$ is defined as
\begin{eqnarray}
\tau (r,z) = \int_0^r \left( \kappa_{\mathrm{a}} +  
                       \kappa_{\mathrm{s}}    \right)  \rho \left( r',\frac{zr'}{r} \right) \sqrt{1 + \left( \frac{z}{r} \right)^2} dr'.
\end{eqnarray}
The conservation equations in the disk then read
\begin{eqnarray}
&{}&\frac{\partial E_{\nu,\mathrm{disk}}}{\partial t} + \nabla \cdot \boldsymbol{F}_{\nu,\mathrm{disk}}=\nonumber\\
&{}&\quad\quad\quad \kappa_{\mathrm{a}} \rho \left[ 4\pi B_\nu(T) - cE_{\nu,\mathrm{disk}}  \right]  
	+  \kappa_\mathrm{s} \rho cE_{\nu,\mathrm{star}},  \label{eqenudisk} \\
&{}&\frac{\partial \boldsymbol{F}_{\nu,\mathrm{disk}}}{\partial t} + c^2 \nabla \cdot \mathcal{P}_{\nu,\mathrm{disk}}  = 
   -(\kappa_{\mathrm{a}}  +  \kappa_{\mathrm{s}}) \rho c \boldsymbol{F}_{\nu,\mathrm{disk}}.
\end{eqnarray}
Assuming the disk to be a static structure, 
the time evolution term for both $E_{\nu,\mathrm{disk}}$ and $\boldsymbol{F}_{\nu,\mathrm{disk}}$ equal zero. 
The mean intensity $J_\nu$ is defined as
\begin{eqnarray}
J_\nu = \frac{cE_\nu}{4\pi},
\end{eqnarray} 
which is consistent with equation (\ref{equationsourcefunction}).
\par
The radiative transfer and hydrostatic equilibrium are calculated in tandem 
to obtain self-consistent $\rho$, $T$, and $J_\nu$ distributions in the disk.
The dust temperature $T$ at each location $(r,z)$ satisfies the condition for thermal equilibrium,
\begin{eqnarray}
\int \kappa_{\mathrm{a}} \left[ E_\nu -  4\pi B_\nu(T)   \right] d\nu = 0.
\end{eqnarray}
Following \citet{muto15}, we assume that the gas and dust are
well mixed in the vertical direction and use the constant gas-to-dust ratio of
$100$ throughout the disk.
The vertical density distribution of the disk is in equilibrium between 
the gas pressure gradient and the $z$-component of the star gravity,
\begin{eqnarray}
\frac{dP}{dz}=\frac{GM_\mathrm{star}z}{\left( r^2 + z^2  \right) ^{3/2}} \rho_\mathrm{gas},
\end{eqnarray}
where $G$, $M_\mathrm{star}$, and $\rho_\mathrm{gas}$
denote the gravitational constant, the mass of the star,
and the density of the gas, respectively. 
The gas pressure $P$ follows
\begin{eqnarray}
P=\frac{k_\mathrm{B}}{\mu m_\mathrm{H}} \rho_\mathrm{gas} T,
\end{eqnarray}
where $k_\mathrm{B}$ denotes the Boltzmann constant and $m_\mathrm{H}$ the mass of hydrogen atom.
The mean molecular weight is assumed to be $\mu=2.339$.


\begin{thebibliography}{}
\bibitem[Aikawa \& Nomura(2006)]{aikawa06} Aikawa, Y., \& Nomura, H.\ 2006, \apj, 642, 1152 
\bibitem[Anders \& Grevesse(1989)]{anders89} Anders, E., \& Grevesse, N.\ 1989, \gca, 53, 197
\bibitem[Andrews et al.(2011)]{andrews11} Andrews, S.~M., Wilner, D.~J., Espaillat, C., et al.\ 2011, \apj, 732, 42 
\bibitem[Baruteau \& Zhu(2016)]{baruteau16} Baruteau, C., \& Zhu, Z.\ 2016, \mnras, 458, 3927  
\bibitem[Biller et al.(2012)]{biller12} Biller, B., Lacour, S., Juh{\'a}sz, A., et al.\ 2012, \apjl, 753, L38 
\bibitem[Birnstiel et al.(2013)]{birnstiel13} Birnstiel, T., Dullemond, C.~P., \& Pinilla, P.\ 2013, \aap, 550, L8
\bibitem[Bohren \& Huffman(1983)]{bh83} Bohren, C.~F., \& Huffman, D.~R.\ 1983, 
Absorption and Scattering of Light by Small Particles (New York: Wiley)
\bibitem[Brown et al.(2009)]{brown09} Brown, J.~M., Blake, G.~A., Qi, C., et al.\ 2009, \apj, 704, 496 
\bibitem[Calvet et al.(2002)]{calvet02} Calvet, N., D'Alessio, P., Hartmann, L., et al.\ 2002, \apj, 568, 1008
\bibitem[Calvet et al.(2005)]{calvet05} Calvet, N., D'Alessio, P., Watson, D.~M., et al.\ 2005, \apjl, 630, L185
\bibitem[Casassus et al.(2013)]{casassus13} Casassus, S., van der Plas, G., M, S.~P., et al.\ 2013, \nat, 493, 191 
\bibitem[Casassus et al.(2015)]{casassus15} Casassus, S., Wright, C.~M., Marino, S., et al.\ 2015, \apj, 812, 126 
\bibitem[Chiang \& Goldreich(1997)]{chiang97} Chiang, E.~I., \& Goldreich, P.\ 1997, \apj, 490, 368 
\bibitem[Close et al.(2014)]{close14} Close, L.~M., Follette, K.~B., Males, J.~R., et al.\ 2014, \apjl, 781, L30  
\bibitem[Espaillat et al.(2014)]{espaillat14} Espaillat, C., Muzerolle, J., Najita, J., et al.\ 2014, Protostars and Planets VI, 497 
\bibitem[Fujiwara et al.(2006)]{fujiwara06} Fujiwara, H., Honda, M., Kataza, H., et al.\ 2006, \apjl, 644, L133 
\bibitem[Fukagawa et al.(2006)]{fukagawa06} Fukagawa, M., Tamura, M., Itoh, Y., et al.\ 2006, \apjl, 636, L153 
\bibitem[Fukagawa et al.(2013)]{fukagawa13} Fukagawa, M., Tsukagoshi, T., Momose, M., et al.\ 2013, \pasj, 65, L14 
\bibitem[Grady et al.(2015)]{grady15} Grady, C., Fukagawa, M., Maruta, Y., et al.\ 2015, \apss, 355, 253 
\bibitem[Gonz{\'a}lez et al.(2007)]{gonzalez07} Gonz{\'a}lez, M., Audit, E., \& Huynh, P.\ 2007, \aap, 464, 429 
\bibitem[Isella et al.(2013)]{isella13} Isella, A., P{\'e}rez, L.~M., Carpenter, J.~M., et al.\ 2013, \apj, 775, 30 
\bibitem[Kanno et al.(2013)]{kanno13} Kanno, Y., Harada, T., \& Hanawa, T.\ 2013, \pasj, 65, 72 
\bibitem[Kataoka et al.(2015)]{kataoka15} Kataoka, A., Muto, T., Momose, M., et al.\ 2015, \apj, 809, 78
\bibitem[Kataoka et al.(2016)]{kataoka16} Kataoka, A., Tsukagoshi, T., Momose, M., et al.\ 2016, \apjl, 831, L12
\bibitem[Lyra \& Lin(2013)]{lyralin13} Lyra, W., \& Lin, M.-K.\ 2013, \apj, 775, 17 
\bibitem[Marino et al.(2015)]{marino15} Marino, S., Perez, S., \& Casassus, S.\ 2015, \apjl, 798, L44 
\bibitem[Mendigut{\'{\i}}a et al.(2014)]{mendigutia14} Mendigut{\'{\i}}a, I., Fairlamb, J., Montesinos, B., et al.\ 2014, \apj, 790, 21 
\bibitem[Mittal \& Chiang(2015)]{mittalchiang15} Mittal, T., \& Chiang, E.\ 2015, \apjl, 798, L25 
\bibitem[Muto et al.(2015)]{muto15} Muto, T., Tsukagoshi, T., Momose, M., et al.\ 2015, \pasj, 67, 122 
\bibitem[Muzerolle et al.(2010)]{muzerolle10} Muzerolle, J., Allen, L.~E., Megeath, S.~T., Hern{\'a}ndez, J., \& Gutermuth, R.~A.\ 2010, \apj, 708, 1107 
\bibitem[P{\'e}rez et al.(2014)]{perez14} P{\'e}rez, L.~M., Isella, A., Carpenter, J.~M., \& Chandler, C.~J.\ 2014, \apjl, 783, L13 
\bibitem[Rodigas et al.(2014)]{rodigas14} Rodigas, T.~J., Follette, K.~B., Weinberger, A., Close, L., \& Hines, D.~C.\ 2014, \apjl, 791, L37 
\bibitem[Rybicki \& Lightman(1979)]{rybicki79} Rybicki, G.~B., \& Lightman, A.~P.\ 1979, New York, Wiley-Interscience, 1979.~393 p.,  
\bibitem[Skrutskie et al.(1990)]{skrutskie90} Skrutskie, M.~F., Dutkevitch, D., Strom, S.~E., et al.\ 1990, \aj, 99, 1187 
\bibitem[Strom et al.(1989)]{strom89} Strom, K.~M., Strom, S.~E., Edwards, S., Cabrit, S., \& Skrutskie, M.~F.\ 1989, \aj, 97, 1451 
\bibitem[Tazaki et al.(2016)]{tazaki16} Tazaki, R., Tanaka, H., Okuzumi, S., Kataoka, A., \& Nomura, H.\ 2016, \apj, 823, 70 
\bibitem[van den Ancker et al.(1998)]{vandenancker98} van den Ancker, M.~E., de Winter, D., \& Tjin A Djie, H.~R.~E.\ 1998, \aap, 330, 145 
\bibitem[van der Marel et al.(2013)]{marel13} van der Marel, N., van Dishoeck, E.~F., Bruderer, S., et al.\ 2013, Science, 340, 1199 
\bibitem[van der Marel et al.(2016)]{marel16} van der Marel, N., van Dishoeck, E.~F., Bruderer, S., et al.\ 2016, \aap, 585, A58 
\bibitem[Verhoeff et al.(2011)]{verhoeff11} Verhoeff, A.~P., Min, M., Pantin, E., et al.\ 2011, \aap, 528, A91 
\bibitem[Williams \& Cieza(2011)]{cieza11} Williams, J.~P., \& Cieza, L.~A.\ 2011, \araa, 49, 67 
\bibitem[Zhu \& Stone(2014)]{zhu14} Zhu, Z., \& Stone, J.~M.\ 2014, \apj, 795, 53 
\end{thebibliography}
\end{document}